\documentclass[11pt,a4paper]{article}

\usepackage{jheppub}     
\makeatletter
\def\@fpheader{}
\makeatother
\usepackage{mathtools}
\usepackage{braket}
\usepackage{bbm}         

\usepackage{graphicx}

\usepackage{feynmp-auto}
\usepackage{booktabs}
\usepackage{tabularx}
\usepackage{array}
\usepackage{makecell}
\usepackage[table]{xcolor}

\usepackage{cancel}
\usepackage{indentfirst}
\usepackage{enumitem}
\usepackage{float}
\usepackage{comment}
\usepackage[normalem]{ulem}  

\newcommand{\be}{\begin{equation}}
\newcommand{\ee}{\end{equation}}

\title{\boldmath Broken Weyl Gravity: Dynamical Torsion and Generalized Proca}

\author{Lucas Fernández\textendash Sarmiento}
\author{Irvin Martínez\textendash Rodríguez}

\affiliation{Department of Physics, Carnegie Mellon University, Pittsburgh, PA 15213, USA}

\emailAdd{lucasf@andrew.cmu.edu}
\emailAdd{ifmartin@andrew.cmu.edu}

\abstract{
We develop a symmetry-based construction of gravity from a phase in which Weyl invariance is spontaneously broken. Using the coset formalism, the dilaton acts as a Stückelberg field for the gauged scale symmetry, giving the Weyl gauge field a mass of order $\mathcal{O}(M_{\rm Pl})$. The resulting gauge–gravity interactions fall within the ghost-free generalized Proca framework. We further show that the Weyl field can be exchanged for a propagating vector component of torsion, providing a natural realization of dynamical vector torsion. Our construction organizes all operators systematically, clarifies the absence of Wess–Zumino terms in $d=4$, and recovers General Relativity in the infrared with controlled higher-derivative corrections.
}

\begin{document} 
\maketitle
\flushbottom

\section{Introduction}


Gauge symmetries and spontaneous symmetry breaking are central concepts in modern theoretical physics. Gauge theories provide the natural language for describing the fundamental interactions: the standard model of particle physics is built upon local gauge invariances, with the dynamics of the electroweak and strong interactions dictated by their associated gauge fields. General relativity can also be understood as a gauge theory by gauging the Poincaré group, with the vierbein playing the role of the gauge field of local translations and the spin connection serving as the gauge connection of the local Lorentz group \cite{Ivanov:1981wn,Delacretaz:2014oxa}, in close analogy with Yang–Mills theories. Just as spontaneous symmetry breaking determines the physical phases of gauge theories such as the electroweak theory or QCD, it may also offer a unifying perspective on the gauge formulation of gravity, where different realizations of symmetry breaking correspond to distinct gravitational phases. 

Scale invariance (symmetry under global rescalings of all lengths) and its gauged counterpart, Weyl invariance (local rescaling of the metric), have played an essential role in the advancement of quantum field theory and gravity. In this work, we study the form and features of theories where this symmetry is spontaneously broken.  Scale invariance arises naturally at renormalization-group fixed points and represents the universal hallmark of critical phenomena, governing systems ranging from QCD at high temperature to quantum phase transitions in condensed matter. Conformal and scale symmetries have been pivotal in addressing fundamental problems such as proposing solutions to the hierarchy problem~\cite{Alexander-Nunneley:2010tyr, Salvio:2020axm, Shaposhnikov:2018nnm}, bootstrapping theories~\cite{Poland2018TheCB, DiFrancesco:1997nkconformalyellow}, and exploring novel gravitational phenomena~\cite{Johansson:2017srf, Iorio:1996ad, Quiros:2014hua, 1985arkady}.

In two dimensions, scale invariance is automatically enhanced to full conformal symmetry under mild assumptions such as unitarity and locality~\cite{Polchinski:1987dy,Luty:2012ww}. However, a general proof for conditions of this enhancement in higher dimensions remains elusive. While most known examples do flow to conformality, explicit counter-examples exist~\cite{Nakayama2016}, and theoretical arguments suggest the two symmetries may part ways in non-trivial corners of parameter space~\cite{nakayama2014scale}. On the other hand, it can be shown that Weyl invariant theories are conformal when the space-time is taken to be flat \cite{nakayama2014scale}. Because not every scale-invariant theory enjoys conformal enhancement~\cite{Riva:2005gd,El-Showk:2011xbs,gravitycftFarnsworth:2021zgj}, in this work we develop a systematic low-energy description of the gauge gravitational theory with broken scale symmetry.

The difference between Weyl invariance and conformal transformations is that Weyl transformations act on the metric $g_{\mu \nu}$ and field $\Phi$ while leaving the coordinates unchanged, 
\begin{align}
    g'_{\mu\nu}(x) =& \, e^{2\omega(x)} g_{\mu\nu}(x),\\
    \Phi'(x) =& \, e^{\Delta\omega(x)} \Phi(x)
\end{align}
where $\Delta$ is the conformal weight of the field $\Phi$, with $\Delta=2$ for the metric, and $\Delta=-\frac{d-2}{2}$ for scalars in $d$-dimensions. This is in juxtaposition with the action of conformal symmetries on fields, which on top of a rescaling by a conformal weight, the transformations act actively on the coordinates,  thus transforming physically inequivalent configurations, while Weyl symmetry tells us that theories with rescaled metric are physically equivalent.

Exact scale and Weyl invariances must eventually be broken in theories describing our universe, either spontaneously or explicitly through anomalies or boundary conditions~\cite{Deser:1993yx, Capper:1974ic, Duff:1993wm, Henningson:1998gx, Bilal:2008qx}, giving rise to a dilaton or conformal anomaly. The spontaneous breaking of global scale invariance produces the dilaton, a  Nambu-Goldstone mode which transforms non-linearly,
\begin{equation}
  \varphi(x)\;\longrightarrow\; \varphi'(x)=\varphi( x)-\lambda(x).
\end{equation}

\noindent Dilaton dynamics are ubiquitous in the current theoretical landscape: in string theory the expectation value of $\varphi$ sets the string coupling; in scalar-tensor gravity it underpins Brans-Dicke inflation~\cite{branskdicke.124.925}; and in holography it captures departures from conformality. In models with extra dimensions the same field reappears as the \emph{radion}, controlling the compactification radius of Kaluza-Klein spaces~\cite{Bailint1987} or the brane separation in Randall-Sundrum type models in extra dimensions ~\cite{Randall1999, Goldberger:1999uk,Rattazzi2001,Arkani-Hamed2000}. 



In this work, we derive the most general effective field theory for gauged Poincaré with spontaneously broken Weyl and show that it naturally organizes itself into an Einstein-Cartan framework with a healthy Proca field \cite{Ghilencea:2019jux} that can be exchanged for a propagating vector-torsion mode. We employ the coset construction ~\cite{Coleman:1969,Callan:1969sn} to ensure that all interaction terms transform appropriately under both broken and unbroken symmetries. We start by gauging the Poincaré group \cite{Ivanov:1981wn} with vierbein $E_{\mu}^a$ and spin connection $\omega_{\mu}^{ab}$ , and gauging scale symmetry \cite{Iorio:1996ad} with gauge connection $A_{\mu}$.

For the Weyl-unbroken phase, a prerequisite to obtaining Weyl in curved space-time is that the flat space-time theory can be enhanced to be conformal, that is, if there exists a CCJ\cite{Callan:1970ze} improvement of the stress energy tensor to make it traceless.  Their work identified a specific gauge field combination that promotes flat-space Weyl invariance to curved space-time geometries. While this combination initially appeared ad-hoc, Karananas and Monin~\cite{Karananas2015WeylConstruction} later showed that it emerges naturally within the coset construction framework by considering the most general Weyl invariant theories and imposing a particular constraint. Our work extends this approach by considering non-linearly realized dilation symmetry, in contrast to previous studies that focused on unbroken Weyl invariance. 

Once Weyl symmetry is spontaneously broken, the gauge field $A_{\mu}$ becomes a massive Proca boson via the Stückelberg mechanism~\cite{PhysRevLett.13.508,Ghilencea:2018dqd}. 
Working in unitary gauge eliminates explicit dilaton dependence from the action, yielding a theory containing a massive vector boson coupled to a massless graviton. Given that the mass term comes from the Ricci scalar modified by the gauge field $A_{\mu}$, the gauge field acquires a mass of the order of the Planck mass $M_{\rm Pl}$. 

Remarkably, at leading order in the EFT, the gauge–gravity couplings produced by our construction fall within the generalized Proca family. These operators extend the Proca action with carefully built derivative self interactions that keep the vector at three propagating modes and avoid Ostrogradsky ghosts, in close analogy with Galileon theories \cite{HeisenbergProca:2017mzp,Heisenberg:2016eld}. In the coset construction for spontaneously broken Weyl symmetry, this structure emerges automatically once we assume the usual Maxwell kinetic term and take the light spectrum to be the graviton and a massive vector. In particular, we do not  include higher curvature terms, as these introduce additional degrees of freedom in the gravity sector, which we do not include at this order.
 No ad-hoc restrictions on curvature–gauge terms are imposed to obtain the generalized Proca interactions.

\begin{figure}
    \centering
    \includegraphics[width=0.65\linewidth]{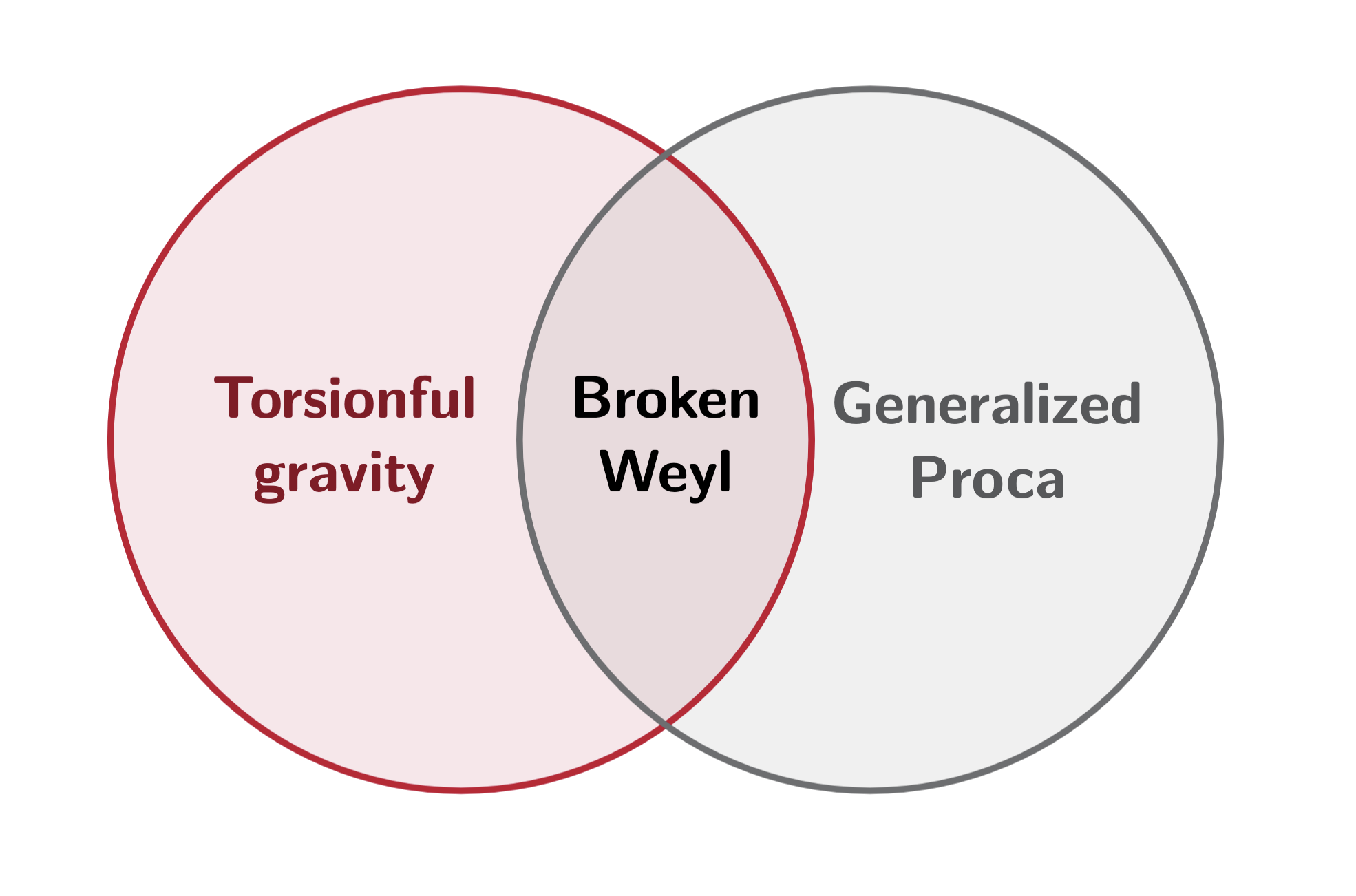}
    \caption{Healthy torsionful gravity with a propagating vector torsion is in one-to-one correspondence with Broken Weyl gravity. Unlike many torsionful models that are plagued by pathologies, Broken Weyl is a subset of the generalized Proca class and is ghost-free by construction.}
\label{fig1}

\end{figure}

One of the main results of this work is the resurrection of torsionful gravity by showing its equivalence to theories in which Weyl symmetry is spontaneously broken. In \cite{Barker:2023fem,Barker:2024goa}, it was shown that the vector component of torsion can be rendered ghost-free by introducing appropriate Lagrange multipliers. Our work shows that this mechanism emerges naturally from the coset construction of spontaneously broken Weyl symmetry, thereby providing a symmetry-driven realization of a healthy, dynamical torsion vector, as depicted in Fig~\ref{fig1}.%





Massive Proca vectors offer fertile ground for phenomenology: massive vectors have been invoked for inflation~\cite{inflationmassive1PhysRevD.40.967,inflationChiba:2008eh,inflationKanno:2008gn,Graham:2015rva,Ghilencea:2020rxc,Cai:2021png}, for late-time acceleration~\cite{Tasinato:2014eka,Armendariz-Picon:2004say}, and as dark-matter candidates~\cite{Chung:1998ua, Baek:2012se,Arcadi:2019lka,Arcadi:2020jqf,Belyaev:2016icc,Farzan:2012hh, Hancock:2025ois, Fernandes:2025lon}. Although the very heavy mass we derive for the Proca field restricts the phenomenological implications that can be possibly tested, our construction provides a unified description of gauge gravitational theories based on symmetry principles from which we can understand our gravitational theory as being in a broken phase. In this phase, we recover the theory of general relativity with modifications that could potentially have observable imprints.

Our results are derived in four dimensions, but using the building blocks, it is straightforward to extend this to n-dimensional theories by simply writing all appropriately contracted $n$-forms. The paper is organized as follows. Section \ref{sec:cosetconstructio} introduces the coset construction which is essential for our framework. Section \ref{sec:BrokenWeylgravity} derives the building blocks of gravity with Broken Weyl symmetry, contrasts them with unbroken Weyl, and constructs the invariant operators that define the effective action. Section \ref{sec:Wess-Zumino} shows that there are no Wess-Zumino terms. Section \ref{sec:Generalized Proca} demonstrates that the leading resulting effective action lies in the generalized Proca class. Section \ref{sec:Torsion} shows that the Proca field can be seen as dynamical Vector Torsion. Section \ref{sec:approximatesymmetries} discusses approximate symmetries. Finally, Section \ref{sec:discussion} concludes with phenomenological implications and directions for future work. The notation conventions can be found in the Appendix \ref{app:conventions}.

\section{Coset construction}
\label{sec:cosetconstructio}

Symmetry breaking played a key role when the first massless excitations were found in the context of superconductors \cite{PhysRev.122.345,PhysRev.124.246}, which would later be generalized to the ubiquitous Goldstone theorem \cite{Goldstone:1961eq,PhysRev.127.965}. Albeit its initial motivation in the context of global symmetries, it would not take long until it was noticed that symmetry breaking also played a role in gauge theories \cite{PhysRev.130.439,PhysRevLett.13.321}, culminating in the Higgs mechanism \cite{PhysRevLett.13.508,PhysRevLett.13.585} that would explain the electroweak spectrum through electroweak symmetry breaking \cite{Glashow:1961tr,Weinberg:1967tq,Salam:1968rm}. 


The coset construction was initially introduced as a way to consistently write down theories consistent with certain symmetry-breaking patterns. One of its earliest uses was to understand the structure of chiral symmetry breaking, and the formalism was introduced by Callan, Coleman, Wess $\&$ Zumino \cite{Coleman:1969,Callan:1969sn,Bardeen:1969ra,PhysRev.184.1750}. It was later extended to include space-time symmetries \cite{salamcft.184.1760,ogievetsky1974nonlinear}, and would play an important role in subsequent formal developments, such as CFTs \cite{Goddard:1984vk}, or String theory \cite{Metsaev:1998it}. It has recently seen a surge in popularity among effective field theories from modifications of gravity, to condensed matter systems \cite{Armendariz-Picon:2010nss,Goon2012GalileonsTerms,equivalenceofconstructionsCreminelli:2013fxa,Delacretaz:2014oxa,Nicolis:2013lma,Goon:2014ika,Karananas2015WeylConstruction,Goon:2014paa,conformalcosetPhysRevD.98.025006,gaugedgal,Akyuz:2023lsm}. Here, we apply this construction to a theory where a scale invariant theory is gauged, and the scale invariance is broken. We will proceed to lay out the general process for building theories using the coset construction, and then apply it to our construction.

We will focus on space-time symmetry breaking as it is most suitable, but general reviews can be found elsewhere \cite{Weinberg:1996kr,Penco:2020kvy}. 
We consider a theory with symmetry group $G$, which has a ground state that spontaneously breaks it to a subgroup $H$. Let us assume the translations $P_a$ are unbroken, so that we can separate the generators into those which are unbroken, $P_a, T_A$, and those which are broken, $Z_\alpha$, with $A\in{1,\dots,\dim H}$ and $\alpha\in{1,\dots,\dim(G/H)}$. Together, $P_a$, $T_A$, and $Z_\alpha$ form a basis of the Lie algebra $\mathfrak{g}$ of the full symmetry group $G$. We can then write the \textit{coset parametrization} as
\begin{equation}
g = e^{i x_a P^a} e^{i \pi_\alpha(x) Z^\alpha},
\end{equation}
where $x_a$ is the space-time position at which we are evaluating the fields, and $\pi_\alpha(x)$ are the Nambu–Goldstone (NG) fields.

The appearance of the factor $e^{i x_a P^a}$ has two motivations. First, translations act on the coordinates non‑linearly, $x^\mu \to x^\mu + a^\mu$, so it is natural to group them with the broken symmetry generators regardless of whether they are broken. This is because translations shift the coordinates; hence their action on the coordinate label is affine rather than linear like the other unbroken symmetries. Second, the factor can be interpreted as translating the origin to the space-time point at which we evaluate the NG fields. 

A general symmetry transformation $g \to \tilde{g}, g$ can always be decomposed into an element independent of the unbroken generators $T_A$ and an element of the coset form:
\begin{equation}
\tilde{g} \cdot g(x, \pi) = g(\tilde{x}, \tilde{\pi}) \cdot h(\pi, \tilde{g}),
\end{equation}
where $h \in H$. This defines a non‑linear transformation of the Goldstone fields, $\tilde{g}: \pi(x) \to \tilde{\pi}(\tilde{x})$, which becomes linear in the special case $\tilde{g} \subset H$ \cite{Goon:2014ika}.

With this we can define the algebra‑valued Maurer–Cartan form
\begin{equation}
g^{-1} d g = E^a P_a + \omega^\alpha Z_\alpha + \omega^A T_A,
\end{equation}
and the fact that $g^{-1} d g \in \mathfrak{g}$ ensures this decomposition is always possible.
Under a symmetry transformation $\tilde{g}$, we obtain the following transformations: 
\begin{equation}
\tilde{g}: \begin{cases}\omega_\alpha & \longmapsto h(x) \omega_\alpha h^{-1}(x) \\ \omega_A & \longmapsto h(x)\left(\omega_A+\mathrm{d}\right) h^{-1}(x)\end{cases}
\end{equation}

The translation coefficients $E_\mu^{\;\;a}$ can be interpreted as vierbeins given their transformation properties \cite{gaugedgal}, and we can define a metric as $g_{\mu\nu}=\eta_{ab}E_\mu^{\;\;a}E_{\nu}^{\;\;b}$. The one-forms $\omega^\alpha$ accompanying the broken generators transform covariantly under all symmetries, and are often recast into $\nabla_a \pi^\alpha\equiv{(E^{-1})}^\mu_{\;\;a} \omega_\mu^{\;\;\alpha}$ and are referred to as the coset covariant derivatives, as they can be expanded as $\nabla_a \pi^\alpha=\partial_a \pi^\alpha+\mathcal{O}(\pi^2)$. Finally, the fields $\omega^A$ are referred to as coset connection, due to their transformation properties which make them act as a $H$-connection and can be used to construct covariant derivatives. These are the building blocks, as we will use them to build the effective action realizing the broken symmetries non-linearly, while keeping linearity in the unbroken ones by simply contracting their indices appropriately. The coset construction can be elegantly extended to include gauge symmetries \cite{Delacretaz:2014oxa}. Consider a gauged subgroup $G_g$ of the symmetry group G, with generators $V^I$  by introducing gauge fields as connections into the Maurer-Cartan form as 
\begin{equation}
    g^{-1}(d+A_IV^I)g
\end{equation}
which is invariant under local gauge transformations 
\begin{equation}
g \rightarrow \tilde{g}(x) g, \quad A^I V_I \rightarrow  \tilde{g}(x) A^I V_I  \tilde{g}^{-1}(x)+ \tilde{g}(x) \mathrm{d}  \tilde{g}^{-1}(x)
\end{equation}
for $g\in G_g$. For a more in-depth discussion, see \cite{gaugedgal,Delacretaz:2014oxa}. Our final ingredient is a kinetic term for the gauge fields, which can be obtained through the Maurer-Cartan two-form 
\begin{equation}
    g^{-1}(d+A_IV^I)^2g
\end{equation}
which will yield the field strengths, and curvature tensors for our connections. 

\subsection{Inverse Higgs constraints}

When space-time symmetries are broken, the number of Goldstone
fields introduced by the coset construction often overcounts the number of
truly independent Nambu-Goldstone (NG) modes.
A systematic way to discard the redundant ones is given by the
\textit{Inverse Higgs Constraint} (IHC) \cite{Ivanov:1975zq}.
Specifically, let \(X_{(1)}\) and \(X_{(2)}\) be two broken generators
such that the commutator of an unbroken translation with the former
closes onto the latter:
\begin{equation}
  [P_\mu , X_{(1)}] = i f_{\mu (1)}{}^{(2)}\, X_{(2)} \; .
\end{equation}
Because the covariant derivative \(\nabla_\mu \pi^{(2)}\) already
transforms as \(X_{(2)}\), one may impose the algebraic condition
\begin{equation}
  f_{\mu (1)}{}^{(2)}\,\nabla_\mu \pi^{(2)} = 0 \; ,
\end{equation}
which can be solved locally to express the field \(\pi^{(1)}\) in
terms of derivatives of \(\pi^{(2)}\). This eliminates the redundant
Goldstone without breaking any symmetry, yielding the minimal field
content for the chosen breaking pattern.

In the scale-invariant example at hand, the only additional broken
generator is the dilation \(D\). Its algebra with translations reads
\begin{equation}
  [P_\mu , D] = - i P_\mu \; ,
\end{equation}
and therefore closes onto the \emph{unbroken} generator \(P_\mu\)
rather than another broken one. Thus, no IHC is needed in this case and
the dilaton remains an independent NG mode. The discussion is
included here for completeness.

\section{Breaking Weyl}
\label{sec:BrokenWeylgravity}
The theory of General Relativity (GR) can be recovered in the coset formulation by gauging the Poincaré group \(ISO(3,1)\) and treating space-time translations as non-linearly realized symmetries \cite{Ivanov:1981wn,Delacretaz:2014oxa}. Adding the single dilation generator \(D = -i\,x^{\mu}\partial_{\mu}\) \cite{Qualls:2015qjbcftlecture} introduces one new relation, \([P_{\mu},D] = -iP_{\mu}\), and enlarges the symmetry \(ISO(3,1)\rtimes\mathbb R\). In this work we consider the case in which the latter is gauged and spontaneously broken down to Poincaré. A key motivation for gauging dilatations alongside Poincaré is to probe the novel features of a $U(1)$-like symmetry that mixes with space-time generators. As in an internal $U(1)$, the associated connection is Higgsed, but the space-time mixing introduces new subtleties. In particular, the geometry becomes Weyl rather than purely Riemannian, and the Weyl gauge field acquires a mass of order $M_{\rm Pl}$ as we shall see.
 By parameterizing the coset, $ISO (3,1)\rtimes \mathbb{R} / SO(3,1)$,  the coset element reads

\begin{equation}
g = e^{i y^a (x) P_a} e^{i \varphi (x) D},
\label{eq:gravscalardilparam}
\end{equation}

\noindent with $y^a(x)$ the local inertial coordinates to describe a curved background \cite{Delacretaz:2014oxa} and $\varphi (x)$ the scalar field parameterizing local dilations.

Then, gauging the whole group, with $A$ the gauge field for dilations,  the covariant Maurer-Cartan form reads 
\begin{align}
   g^{-1} \mathcal{D} g &= g^{-1} \left( \mathrm{d} + i \tilde{E}^{a} P_a + \frac{i}{2} \omega^{ab} J_{ab} + i \tilde{A} D \right) g\\
&= \mathrm{d} + i\left( \tilde{E}^{a} + \mathrm{d}y^a - \omega^{a}_{\;b} y^b - \tilde{A} y^a \right) e^{\varphi} P_a + \frac{i}{2} \omega^{ab} J_{ab} + i( \mathrm{d}\varphi + \tilde{A}) D\label{buildingblockfleshedout}\\
&= \mathrm{d} + i\,E^{a}\,P_a   + \frac{i}{2}\,\omega^{ab}\,J_{ab} + i\,\nabla\varphi\,D.\label{blocks}
\end{align}


\noindent The $E_\mu^{\;a}$ is identified with the vierbein such that $E_\mu^{\;a}E^{\mu b}=\eta^{ab}$ and $E_\mu^{\;a}E_{\nu a}=g_{\mu\nu}$\footnote{We can identify $ \bar{E}^a = \tilde{E}^{a} + \mathrm{d}y^a - \omega^{a}_{\;b} y^b  $  with the vierbein of GR, now shifted by the Weyl connection through $ A y^a$ and dilated, as one would expect.}.  The covariant spin connection $\omega^{ab} $ has yet to be solved in terms of the vierbein as shown below. 
The new degree of freedom, $\nabla \varphi = \mathrm{d}\varphi + A$, is the covariant derivative of the Goldstone boson from the breaking of dilations, with $\varphi$ now playing the role of a Stückelberg field: under local Weyl rescaling we find

\begin{subequations}\label{eq:Weyl-transforms}
\begin{align}
\varphi &\;\to\; \varphi-\lambda, &
\tilde{A} &\;\to\; \tilde{A}+\mathrm d\lambda,
\label{eq:Weyl-Stueckelberg}\\[4pt]
\omega_{ab} &\;\to\; \omega_{ab}, &
\tilde{E}^{a} &\;\to\; e^{\lambda} \tilde{E}^{a}, &
y^{a} &\;\to\; e^{\lambda}y^{a} .
\label{eq:Weyl-geometry}
\end{align}
\end{subequations}
\noindent with the geometric fields transforming as expected under a Weyl transformation. A short note is in order regarding the transformation of $y^a$. The space-time coordinates $x^{\mu}$ are inert labels and do not transform under the Weyl gauge symmetry. In contrast, $y^a(x)$ are dynamical Goldstone fields. It so happens that in flat space-time one can choose a gauge in such a way as to identify one with the other via $y^a=\delta^a_{\;\mu}x^\mu$, but in curved space-time they have inherently different interpretations. Because they correspond to broken translations, the $y^a$ have a conformal weight 1. Hence $y^a \to e^{\lambda}y^a$.

Choosing to work in unitary gauge, \(\lambda=\varphi\),  removes the local Weyl redundancy:  the vector shifts to $A'=\tilde{A}+d\varphi$, while the scalar shifts away, $\varphi'=0$. Therefore, the building blocks do not have an explicit dependence on $\varphi$. The gauge field $A'$  eats the Goldstone of broken dilations via the Stückelberg mechanism. Since the explicit dependence on $\varphi$ is washed away, we simply denote $\nabla\varphi \equiv A$.

The torsion  $T^a$ and curvature $R^{ab}$  are encoded in the Maurer-Cartan 2-form,

\begin{align}
    &g^{-1}\Bigl( \text{d} + i\,\tilde{E}^{a}\,P_a  + \frac{i}{2}\,\omega^{ab}\,J_{ab} + i\,\tilde{A}\,D \Bigr)^2g\\
    &\quad \quad   = (\text{d}E^a+E^b\wedge{\omega^a}_b+E^a\wedge A)P_a+(\text{d}\omega^{ab}-\omega^{ac}\wedge \omega_{c}^{\;\;b})J_{ab}+ \mathrm{d}A D\\
    &\quad \quad  = T^a P_a+ R^{ab}J_{ab}+ F D.
\end{align}

Note that the building blocks in the second and third lines are those from the coset construction \ref{blocks}, and already transform appropriately under our coset. 
\noindent To derive these results, we used a choice of the Poincaré algebra often used in differential geometry, leading to a slightly different form of  $T^a$ and  $R^{ab}$. We can easily make the replacement $\omega\rightarrow-\omega,\quad R\rightarrow -R$ to recover the usual equations,

\begin{align}
\begin{split}
    &T^a= \text{d}E^a-E^b\wedge{\omega^a}_b+E^a\wedge A,\\&
    R^{ab}= \text{d}\omega^{ab}+\omega^{ac}\wedge \omega_{c}^{\;\;b},\\&
    F= \mathrm{d}A.
\end{split}
\end{align}

\noindent The 2-form  $T^a $ contains the GR torsion and an additional piece that shifts it by $ E^a \wedge A$. The gauge field strength $F$ is the usual Maxwell two-form.  The Riemann 2-form $R^{ab} $ has the same structure as the one for GR, but now the spin connection, when solved in terms of the vierbein by imposing that the torsion tensor vanishes, will be shifted by $A$.



Setting  $T^a = 0$ is a covariant condition that we may impose for reasons that will become clearer soon. This determines the spin connection  $\omega^{ab}$ in terms of the vierbein and the Weyl gauge field \cite{Karananas2015WeylConstruction},

\begin{equation}
\omega_\mu{}^{a b}=\bar{\omega}_\mu{}^{a b}(E)+2 \left(E^{-1}\right)^{\rho[a} E_{\mu}^{\;\;b]}A_\rho{},
\label{soldilatonconnection}
\end{equation}
where 
\begin{equation}
\bar{\omega}_\mu{}^{a b}{ }=2\left(E^{-1}\right)^{\rho[a} \partial_{[\mu} E_{\rho]}{ }^{b]}-\left(E^{-1}\right)^{\rho[a}\left(E^{-1}\right)^{b] \sigma} E_{\mu c} \partial_\rho E_\sigma{ }^c,
\label{eq:spinconnectionGR}
\end{equation}
\noindent is the usual torsion-free GR connection.  
Consequently, the curvature 2-form acquires explicit $A_{}$ dependence,  shifting the GR spin connection by

\begin{align}
    \delta \omega_\mu^{ab}=I_{\mu \nu}^{ab} A^\nu, \quad I_{\mu \nu}^{ab}=2E_{[\mu}^a E_{\nu]}^b.
\end{align}
Then, the curvature 2-form  \cite{Karananas2015WeylConstruction}
\begin{equation}
    R_{\mu \nu}^{ab}=\bar{R}_{\mu \nu}^{ab}+\delta R_{\mu \nu}^{ab},\label{curvature_solution}
\end{equation}
with
\begin{equation}
    \frac12\bar{R}_{\mu \nu}^{ab}=\partial_{[\mu }\bar{\omega}_{\nu]}^{ab}-\bar{\omega}_{[\mu |c|}^a \bar{\omega}_{\nu]}^{c b},
\end{equation}
and
\begin{align}
\begin{split}
\frac12\delta R_{\mu \nu}^{ab}  =& I_{[\nu |\lambda|}^{ab} \nabla_{\mu]} A^\lambda-E_{[\mu}^a E_{\nu]}^b A^2 -2E_{[\nu}^{\;[a} A^{b]}A_{\mu]}\end{split}
\end{align}
where barred quantity corresponds to the GR solution and $A^a\equiv E_\mu^{\;a}A^\mu$.\footnote{This expression differs by overall signs from \cite{Karananas2015WeylConstruction}, which adopts the opposite sign in the torsion definition. The two are equivalent up to field redefinitions.} Therefore, we recover all the building blocks of GR, shifted by terms with the gauge field $A$.

In general, we can expand the Levi-Civita connection by adding a contortion tensor linear in the Weyl vector field as 
\begin{align}
   \Gamma_{\mu \nu}^\rho=\bar{\Gamma}_{\mu \nu}^\rho-b_1 A^\rho g_{\mu \nu}+b_2 \delta_{(\mu}^\rho A_{\nu)}+b_3 \delta_{[\mu}^\rho A_{\nu]},
\end{align}
leading to a non-metricity term 
\begin{align}
\nabla_\mu g_{\rho \sigma}=\left(b_3-b_2\right) A_\mu g_{\rho \sigma}+\left(2 b_1-b_2-b_3\right) A_{(\rho} g_{\sigma) \mu}.\\
\end{align}
 In our case, we can calculate 
 \begin{equation}
     \Gamma_{\mu \nu}^\rho=\bar{\Gamma}_{\mu \nu}^\rho +g_{\mu \nu}A^\rho -\delta^{\rho}_{\;\mu}A_\nu,
 \end{equation}
with $b_1=b_2=b_3=-1$, meaning that $\nabla_{\mu} g_{\nu \rho}=0$, leading to generalized Weyl geometries \cite{BeltranJimenez:2015pnp,BeltranJimenez:2016wxw}. This is remarkable for two reasons. Firstly, this means that lengths remain invariant as we parallel transport vectors in these geometries, a common pitfall for many pure Weyl geometries \cite{BeltranJimenez:2016wxw}. Additionally, the non-trivial requirement that $b_3=2b_1-b_2$, avoids propagation of extra degrees of freedom leading to ghost-like instabilities \cite{BeltranJimenez:2016wxw}, ensuring the health of such theory. We will see in section \ref{sec:Generalized Proca} that in fact all leading interactions fall under the generalized Proca \cite{Heisenberg:2014rta} umbrella, ensuring they are not plagued by ghosts.


\subsection{Difference from unbroken Weyl}

We take a moment to understand where this theory diverges from previously studied ones, where the Weyl symmetry was not non-linearly realized (unbroken). In \cite{Iorio:1996ad}, the authors show that one can promote conformal invariance in flat space-time to conformal invariance in curved backgrounds by gauging the dilation symmetry in flat space-time and using the gauge field $A_\mu$ in only the following combination: 

\[
\Theta_{\mu\nu} = \nabla_{\mu} A_{\nu} - A_{\mu} A_{\nu} + \frac{1}{2} g_{\mu\nu} A^{\sigma} A_{\sigma}.
\]
The Weyl variation of the symmetrization of the tensor above is independent of the gauge field, and proportional to the Schouten tensor:
\begin{equation}
    S_{\mu\nu} = \bar{R}_{\mu\nu} - \frac{\bar{R}}{2(n - 1)} g_{\mu\nu},
\end{equation}

\noindent such that 

\begin{equation}
   2\Theta_{\mu\nu} \simeq S_{\mu\nu},
 \label{simtensor}
\end{equation}

\noindent where $\simeq$ refers to the fact that both sides of the equation transform in the same manner.

While this prescription worked to achieve the goal of Weyl invariance, the origin of the $\Theta$ tensor was unclear. In  \cite{Karananas2015WeylConstruction}, the authors consider the coset construction of a gauged and unbroken  dilation, and show that upon applying the constraint 

\begin{align}
    R_{(\mu\nu)} = 0, \quad \text{with} \quad R_{\mu\nu} \equiv R_{\mu\sigma}^{ab} E_b^\sigma E_{\nu a},
\end{align}

\noindent we can recover this result, elucidating the origin of the $\Theta$ tensor, which was previously not well understood. By applying the above constraint to eq. (\ref{curvature_solution}), we get two pieces: one proportional to the Ricci tensor, and one including some terms involving $A_\mu$. Explicitly, we find, 
\begin{equation}
    \bar{R}_{\mu\nu} 
    + \nabla_{\mu} A_{\nu} + \nabla_{\nu} A_{\mu} 
    + g_{\mu\nu} \nabla_{\rho} A^{\rho} 
    - 2 A_{\mu} A_{\nu} 
    + 2 A_{\rho} A^{\rho} g_{\mu\nu} = 0.
\end{equation}
\noindent Tracing this equation, we can find the Ricci scalar,
\begin{equation}
    \bar{R}=-6( A_{\mu  } A^{\mu  } +  \nabla_{\mu  }A^{\mu  }).
\end{equation}

Using these results, one can construct the Schouten tensor and verify eq.~(\ref{simtensor}).\footnote{See Appendix~D for the derivation of the Weyl variation.} The key point is that the combination $X_{\mu\nu}\equiv S_{\mu\nu}-2\Theta_{\mu\nu}$ transforms covariantly (its Weyl weight vanishes). Hence we may consistently impose the covariant constraint $X_{\mu\nu}=0$, independently of the gauge choice at the principal bundle level, before pulling back to any local section.
 Thus, if we build an action with $\Theta$, and $\Theta$ alone (no other $A$ dependence), imposing the covariant constraint $X_{\mu\nu}=0$, allows us to trade all the $A$ dependence for curvature tensors through the Schouten tensor. This idea breaks down in our case for the following reason: $A$ does not transform under Weyl transformations any longer, it is now a Weyl invariant, and a Lorentz tensor. Thus, the identification of the variation of the Schouten tensor with the variation of $\Theta$ no longer holds. 
The antisymmetric part of $\Theta_{\mu\nu}$, upon imposing this constraint, can be shown to be proportional to $F$ in a straightforward manner. 
In \cite{Karananas2015WeylConstruction}, the authors show that upon imposing $R_{\mu \nu}^{ab} E_\nu^{\;b}=0$, the lowest order 
 the action that can be built is 
\begin{equation}
    S=  \int (c_1 F\wedge\star F +c_2 F\wedge F +c_3 \, \mathrm{Tr}( \bar{W} \wedge \star  \bar{W}) + c_4 \, \mathrm{Tr}( \bar{W} \wedge  \bar{W}))
\end{equation}
where $\bar{W}$ is the Weyl tensor of GR, :
\begin{equation}
 \bar{W}_{\mu\nu\rho\sigma}=  \bar{R}_{\mu\nu\rho\sigma}- \frac{2}{n-2}\bigl(  g_{\mu[\rho}\bar{R}_{\sigma]\nu}  - g_{\nu[\rho}\bar{R}_{\sigma]\mu}\bigr)+ \frac{2}{(n-1)(n-2)}\,\bar{R}\,g_{\mu[\rho}g_{\sigma]\nu}\,.
\end{equation}

Once Weyl symmetry is broken, the field $A_\mu$ ceases to behave as a Weyl connection and instead transforms homogeneously as a Lorentz tensor, while the spin connection $\omega_\mu{}^{ab}$ still transforms as a bona fide connection. Consequently, the tensor $\Theta_{\mu\nu}$ no longer transforms in the unbroken case, so the combination $X_{\mu\nu}=S_{\mu\nu}-2\Theta_{\mu\nu}$ is not Weyl-covariant and the constraint $X_{\mu\nu}=0$ cannot be imposed in a gauge covariant way. We therefore proceed to classify all admissible terms without this constraint.

Additionally, Weyl gravity with a Weyl-squared term propagates six degrees of freedom. Expanding around an Einstein background with cosmological constant \( \Lambda \) and in transverse–traceless gauge, the linearised equations read \cite{Lu:2011ks}
\begin{equation}
    \left(\square-\frac{2}{3}\Lambda\right)\left(\square-\frac{4}{3}\Lambda\right) h_{\mu\nu}=0,
\end{equation}
which factorises into a massless Spin-2 graviton and a massive Spin-2 graviton with mass \( M^{2}=\tfrac{2}{3}\Lambda \). This is the partially massless point in \( d=4 \), where the helicity-0 component of the Spin-2 graviton does not propagate. The appearance of ghosts is inevitable in these theories \cite{stellequadraPhysRevD.16.953,Lu:2011ks}, however, as we will see, this is avoided in the spontaneously broken setup, where the propagating modes are a massless Spin-2 graviton and a massive Spin-1 generalized Proca field, both ghost-free. 

\subsection{Bootstrapping the Gravity sector}
We consider the different terms that can be built given our invariants. It will become clear that this task is rather constrained if we want to have any sensible theory and maintain our degrees of freedom unscathed. First, we consider the usual cosmological constant and Einstein-Hilbert terms:
\begin{equation}
\mathcal{L}_{\Lambda_{}} +\mathcal{L}_{\rm EH}= \frac{M_{\rm Pl}^2}{4} \varepsilon_{abcd}(- \frac{1}{6}\Lambda_{cc}  E^a\wedge E^b\wedge E^c\wedge E^d +R^{ab}\wedge E^c \wedge E^d).\label{eq:EHLambdaaction}
\end{equation}
These terms appear in conventional general relativity, with $\Lambda_{\rm cc}$ the cosmological constant. 
In standard formulations with the curvature two-form in four dimensions, the term $R^{ab}\wedge E_a \wedge E_b$, is topological when the torsion vanishes as it is related to the Nieh-Yan 4-form, $N \equiv T^a \wedge T_a-E^{ a} \wedge E^{ b} \wedge R_{a b}=\textnormal{d}\left(E^{a} \wedge T_a\right)$. This topological property also holds with the specific torsion arising in our model of spontaneously broken Weyl symmetry, 
where the $A$ dependence cancels, and the term remains topological.

Therefore, to linear order in the curvature two-forms, only the term in eq.~(\ref{eq:EHLambdaaction}) 
survives, yielding the Einstein–Hilbert action. The key difference here is that our spin connection includes the Weyl gauge field \(A\), and thus the resulting action depends on \(A\). Substituting the solution for the curvature two-form, this term becomes
\begin{equation}
    R \equiv\varepsilon_{abcd}\,R^{ab}\wedge E^c\wedge E^d = E\,\bigl(\bar{R} + 6\,\nabla_\mu A^\mu -6\,A_\mu A^\mu\bigr),\label{eq:R}
\end{equation}
where $\bar{R}$ is the usual Ricci scalar of general relativity, and $E = \operatorname{det} E_{\mu}^a$ . Since the coefficient of $\bar{R}$ scales as \(M_{\mathrm{Pl}}^2\), the Weyl gauge field acquires a mass of order \(M_{\mathrm{Pl}}\), effectively decoupling it at energies well below the Planck scale. With the appropriate normalization of $A_\mu$, its physical mass will be of order $qM_{\rm Pl}$, where $q$ is the Weyl coupling. The term \(6\nabla_\mu A^\mu\) is a total derivative and vanishes under appropriate boundary conditions at infinity. Terms like $\square R$, which can be important for space-times with boundaries and obtained via the Laplacian are also not included here. 


As a consistency check, one may start from the unfixed gauge sector given in equation~(\ref{buildingblockfleshedout}) and set \(A_\mu = 0\). This corresponds to reverting to a theory in which scale invariance is realized non-linearly but not gauged. The resulting Lagrangian density contains the combination $e^{\varphi}\,\bar{R} + 6\square\varphi - 6\,\partial_\mu\varphi\,\partial^\mu\varphi$,
which is precisely the structure expected for a scalar-tensor theory in the Jordan frame obtained by a Weyl rescaling of the metric~\cite{DeFelice:2010aj}.

At second order in $R$, some of the other usual topological suspects remain topological as they do not involve the torsion. The Pontryagin density, 
\begin{equation}
 R^{a}{}_{b} \,\wedge\,R^{b}{}_{a}
  \;=\;
  \textnormal{d}\!\Bigl(
        \omega^{a}{}_{b}\wedge d\omega^{b}{}_{a}
        \;+\;\frac{3}{2}\,
              \omega^{a}{}_{b}\wedge\omega^{b}{}_{c}\wedge\omega^{c}{}_{a}
      \Bigr)\propto E R^{\mu\nu\rho \sigma}\tilde{R}_{\mu\nu\rho\sigma}\equiv P,
\end{equation}
with $\tilde{R}^{\mu\nu\rho\sigma}=\frac12\varepsilon^{\alpha\beta\rho\sigma} R_{\mu\nu\rho\sigma}$ and the Euler density (Gauss-Bonnet) \cite{Nakahara:2003nw}
\begin{equation}
 \mathcal{L}_{\rm GB}=\;
  \frac12\,\varepsilon_{abcd} \; \textnormal{d}\,\Bigl(
      \omega^{ab}\wedge d\omega^{cd}
      \;+\;\frac23\,\omega^{a}{}_{e}\wedge\omega^{eb}\wedge\omega^{cd}
  \Bigr)= \varepsilon_{abcd}R^{ab}\wedge R^{cd}\
  \;\equiv G, 
\end{equation}
are clearly both surface terms, which can be important for theories on manifolds with a boundary. The Pontryagin term, $P$ is odd under parity transformations, and it is the gravitational equivalent of the $\theta$ term in Maxwell, or Yang-Mills theories. The quadratic curvature terms that are topological in the action are ghost-free. Interestingly, as shown in Appendix \ref{Appexpansion}, the Pontryagin term is always accompanied by a gauge theta term in broken Weyl theories. This means that while we can have the gauge theta term without its gravitational counterpart, the reverse is not true. 


Additionally, we may include curvature-squared operators such as 
\begin{equation}
    R^{ab}\wedge \star R_{ab}\propto E\,R_{\mu\nu\rho\sigma}R^{\mu\nu\rho\sigma}. 
\end{equation}

\noindent Using the leading equations of motion, the operators \(R^{2}\) and \(R_{\mu\nu}R^{\mu\nu}\) can be traded for terms proportional to the cosmological constant. In four dimensions one may use the Euler identity to trade the Kretschmann term for Ricci invariants up to a surface term,
\begin{equation}
 R_{\mu\nu\rho\sigma} R^{\mu\nu\rho\sigma}
\;=\;
\left(4\,R_{\mu\nu} R^{\mu\nu}-R^{2}\right)
\;+\mathcal{L}_{GB},
\end{equation}
with \(\mathcal{L}_{GB}\) the Gauss-Bonnet density of the Levi-Civita connection.  Therefore, in broken Weyl theory, similarly to the effective theory of general relativity with higher curvature operators~\cite{Endlich:2017tqa}, these curvature squared terms do not modify the bulk equations at this order, even with a cosmological constant.

\subsection{Pure gauge}

The kinetic term is the usual $-\frac{1}{4q^2}F\wedge \star F$ which yields the standard Maxwell kinetic term in index notation $-\frac{1}{4q^2}F^{\mu\nu}F_{\mu\nu}$, where $q$ is the Weyl coupling. For hygiene, we will set the convention of $q=1$ in what follows, but it is straightforward to  reinstate the couplings if needed. One can also add the usual Pontryagin class of the gauge bundle, $F\wedge F=d(A\wedge F)$, which acts as the theta term. Adding an arbitrary mass term $m^2 A\wedge \star A$ is also allowed by the symmetries, but we note that this will also receive contributions from terms with the gravity curvature two form $R$. To work out the physical mass, one must reintroduce the coupling $q$, so the contribution coming from gravity terms will be of order $qM_{\rm Pl}$.

It is not difficult to show that there are no terms containing three contracted field strengths, or three copies of the gauge field, so the next terms will contain 4 copies of the gauge field, these would be terms such as $(A\wedge\star A)\wedge\star (A\wedge \star A)$ or $F\wedge A\wedge \star (F\wedge A)$, together with squares of the kinetic term, or the $\theta$ term. In general, we denote all such higher order terms as $\mathcal{L} (F,\tilde{F},A)$ 
\color{black}

\subsection{Gauge-gravity interactions}
The leading non-trivial contractions of curvature two-forms with gauge fields, one-forms, or field strength two-forms come from 
\begin{equation}
    R^{ab}\wedge A\wedge \star (A\wedge E_a \wedge E_b),\quad R_{ab} \wedge F \wedge \star(F \wedge E^a \wedge E^b) \quad \varepsilon_{abcd}R^{ab}\wedge F \wedge \star (F\wedge E^c\wedge E^d)
\end{equation}
 and their index form can be found in \ref{Appexpansion}. All other terms including one field strength will either vanish, or be proportional to the kinetic term, and so they can be absorbed into it's coefficient.

\subsection{The Effective Action}
\label{subsec:effective-action}
As we will show in the following section, there are no Wess-Zumino terms, and thus all the invariant terms built above are all the necessary ingredients to build the effective action. This effective theory will be valid in some range between the scale of creation of Weyl gauge fields $m\sim q M_{\rm pl}$, where $q$ is the coupling to the Weyl field, and the scale of symmetry breaking $\Lambda$. Meaning, we require a hierarchy $m\ll\Lambda\ll M_{\rm pl}$. With this in mind, the even-parity action for gauge-gravity with spontaneously broken local dilations in 4d, reads

\begin{align}
\label{eq:action}
\begin{split}
S \;=\; \int \Bigg[
& \frac{M_{\rm Pl}^{2}}{4}\,\varepsilon_{abcd}\!\left(
R^{ab}\wedge E^{c}\wedge E^{d}
- \frac{\Lambda_{\rm cc}}{6}\,E^{a}\wedge E^{b}\wedge E^{c}\wedge E^{d}
\right)
\\
& \;-\; \frac{1}{2}\,F\wedge \star F
\;+\; \frac{m^{2}}{2}\,A\wedge \star A
\;+\; \mathcal{L}(F,\tilde F,A)
\\
& \;+\; \frac{\alpha_{\rm GB}}{\Lambda^2}\,\varepsilon_{abcd}\,R^{ab}\wedge R^{cd}
\;+\; \frac{\alpha_{RA^{2}}}{\Lambda^{2}}\,R^{ab}\wedge A\wedge
\star\!\big(A\wedge E_{a}\wedge E_{b}\big)+\mathcal{O}(\frac{1}{\Lambda^4})
\Bigg] \, .
\end{split}
\end{align}

\noindent   The term $\mathcal{L}(F,\tilde{F},A)$, represents the self-interactions of the gauge built from the gauge field, the field strength, and its dual. Here, $\Lambda_{\rm cc}$ is the cosmological constant, while $\Lambda$ is the symmetry breaking scale. We remind the reader that as well as the allowed $m^2$ term in the Lagrangian, the Weyl field will also receive mass contributions from the Ricci term \ref{eq:R}.   The quadratic curvature term with $\alpha_{RR}$  might or might not be dropped according to whether the topological terms are of interest. 
 

 Coupling to matter that transforms linearly under the unbroken symmetries can be easily added through the covariant derivative 
 \begin{equation}
     \nabla_{a}=\left( E_{a}d+ E_{a} \omega^{cd} J_{cd}\right)
 \end{equation}
where the spin connection $\omega^{ab}$ is the building block of the unbroken generators of the theory. Recalling that $\omega_{}^{ab}$ is the usual general relativistic  spin connection shifted by $A$ as in eq.~(\ref{soldilatonconnection}), then matter fields couple to $A$ through the covariant derivatives. 

As we will show, the action in eq. (\ref{eq:action})  takes a very special form, namely, every parity-even interaction arising from the leading order and next to leading order broken Weyl action falls into the so called generalized Proca theories \cite{HeisenbergProca:2017mzp,Heisenberg:2014rta}. Additionally we can exchange the Proca field completely for propagating vector torsion, giving rise to a completely different perspective on this model as shown in section~\ref{sec:Generalized Proca}.

\section{Wess-Zumino terms, or lack thereof}
\label{sec:Wess-Zumino}

The coset construction allows us to build Lagrangians which are exactly invariant under the proposed symmetries. However, what it does not necessarily do is build all actions invariant under these symmetries. For example, we might have Lagrangians that are invariant up to a total derivative which would be missed in the coset construction, while the action remains invariant. These Lagrangians are generally known as Wess-Zumino (WZ) terms and find a natural interpretation when considering an auxiliary higher dimension \cite{DHoker:1994rdl,Witten:1983tw}.

Consider a $d$-dimensional space-time manifold $\mathcal{N}$. A WZ term in an action on $\mathcal{N}$ is typically constructed from a $d$-form $\alpha$. The defining characteristic of $\alpha$ is that its exterior derivative, $\omega = d\alpha$, is a $(d+1)$-form that is $G$-invariant (where $G$ is the symmetry group of the theory) and represents a non-trivial cocycle in the relative Lie algebra (Chevalley-Eilenberg) cohomology $H^{d+1}(\mathfrak{g},\mathfrak{h})$. This essentially counts $G$-invariant forms on the coset space $G/H$. In the spirit of brevity, we will keep this section succinct and light-hearted, with all the details in the Appendix \ref{CE cohomology}, or in the standard texts for Wess-Zumino terms in the light of cohomology \cite{deAzcarraga:1997qrt,deAzcarraga:1998uy,Goon2012GalileonsTerms}.
This $(d+1)$-form $\omega$ can be thought of as existing on a $(d+1)$-dimensional manifold $\mathcal{M}$ with boundary $\mathcal{N}$, $\partial\mathcal{M} = \mathcal{N}$. The WZ term added to the $d$-dimensional action is then given by 
\begin{equation}
    \mathcal{L}_{WZ} = \int_{\mathcal{N}} \alpha.
\end{equation} 

Using the generalized Stokes' theorem, this can also be written as 

\begin{equation}
    \int_{\mathcal{M}} \omega = \int_{\mathcal{M}} d\alpha = \int_{\mathcal{N}} \alpha.
\end{equation}

For compact homogeneous spaces $G/H$, a classical result ensures that the relative Lie algebra cohomology $H(\mathfrak{g},\mathfrak{h})$ is isomorphic to the de Rham cohomology $H_dR(G/H)$\cite{deAzcarraga:1998uy}. This can then be used to work out an upper bound of WZ terms by finding the dimension of the relevant de Rham cohomology group, in the case of $d=4$, $H_{dR}^5(G/H)$. Nonetheless, in our case we deal with non-compact groups and it does not suffice to borrow the dimensions of the analogous de Rham group, but instead we have to work with the relative cohomology group with no shortcuts.
We take $E_i$ to be the basis of $\mathfrak{g}$, and their duals $\omega^i$ to form a basis of the dual space $\mathfrak{g}^*$ such that $\omega^i(E_j)=\delta^i_{\;j}$. Then we define the structure constants $c_{i j}$  by $\left[E_i, E_j\right]=c_{i j}{ }^k E_k$. The relative cochains are written by wedging the building blocks in all possible ways.

The coboundary operator acts on the forms via the algebra Lie bracket as  
\begin{equation}
 \delta \omega^i=-\frac{1}{2} c_{j k}{ }^i \omega^j \wedge \omega^k .
\end{equation}
 For our purposes, this is rather sparse, given our commutation relations \ref{app:Commutators}. 
We denote our algebra basis as $\{J,P,D\}$, and the Weyl algebra can be used to work out the full action of coboundary operators as
\begin{align}
    \delta D=0,\quad \delta P^a=i(-D \wedge P^a+2 P^b \wedge J^{c a} \eta_{bc}).
\end{align}

The only non-trivial Lorentz scalar 5-form we can build with our ingredients is $\omega_5=\varepsilon_{abcd}D\wedge P^a \wedge P^b \wedge P^c \wedge P^d$. To see why such must be true, we note that we cannot use more than one D, for their exterior product would vanish. Meanwhile, we cannot use more than four $P^\mu$ as otherwise we would be antisymmetrizing five generators with four Lorentz indices, forcing us to repeat at least one, and therefore vanishing.  We can use the cochain rules to show that in the CE complex, we have $\delta \omega_5=0$\footnote{To see this, we note that linear forms in $J$ are relative-CE cohomologous to zero, and once again, the terms with two copies of $D$ would too vanish.} therefore, this form is closed. A similar computation shows that it is exact by computing $\delta \omega_4$ with $ \omega_4= \frac{i}{4} \varepsilon_{abcd}D\wedge P^a \wedge P^b \wedge P^c \wedge P^d$;
\begin{equation}
    \delta \omega_4=i \varepsilon_{abcd} \delta(P^a) \wedge P^b \wedge P^c \wedge P^d=\varepsilon_{abcd}D\wedge P^a \wedge P^b \wedge P^c \wedge P^d=\omega_5
\end{equation}
where the linear $J$ term is set to zero appealing to the fact we are working in the relative CE complex. Hence, all closed relative cochains are exact, and therefore $H^5(\mathfrak{g,\mathfrak{h}})=0$ in our case. With this, we conclude the demonstration of the absence of Wess-Zumino terms in this construction.

\section{Gauge-gravity interactions and broken Weyl as a generalized Proca}
\label{sec:Generalized Proca}
In this section we show that the leading interactions in broken Weyl gravity theories fall under a very restrictive family that ensures ghost-free propagation: generalized Proca. Generalized Proca theories \cite{HeisenbergProca:2017mzp,Heisenberg:2014rta,DeFelice:2016yws,Heisenberg:2018vsk} parallel the developments in other spin sectors, particularly the interacting ghost-free massive graviton (non-linear Fierz--Pauli \cite{Fierz:1939ix}) constructed by de Rham, Gabadadze and Tolley \cite{deRham:2010kj,deRham:2010ik} and proven ghost-free by Hassan and Rosen \cite{Hassan:2011hr,Hassan:2011zd}. Heisenberg found the equivalent model for the Spin-1 case, the generalized Proca actions \cite{Heisenberg:2014rta}.

The key feature is that generalized Proca interactions keep the temporal component $A^0$ non-dynamical, acting as a Lagrange multiplier even in the presence of interactions. This guarantees exactly three propagating degrees of freedom, avoiding the Ostrogradsky instability \cite{ostrogradsky1850memoires}. The structure was motivated by Horndeski's scalar-tensor theories \cite{Horndeski:1974wa,Gleyzes:2014dya}, which showed that second-order equations of motion tightly restrict admissible interactions, a principle later manifested in Galileon theories \cite{Nicolis:2004qq,Nicolis:2008in,Deffayet:2009wt} that emerged as the decoupling limit of massive gravity. For Spin-1 fields, a no-go theorem \cite{Deffayet:2013tca} showed that gauge-invariant vectors cannot have new second-order self-interactions beyond Maxwell. Breaking $U(1)$ gauge symmetry with a mass term circumvents this, allowing additional interactions that preserve the degeneracy condition. 
In \cite{HeisenbergProca:2017mzp}, Heisenberg shows the most general Lagrangian for a generalized Proca theory in both flat and curved space-time. For our purposes, we show the generalized Proca Lagrangian in curved space-time, which can be written as 

\begin{align}
\mathcal{L}_{\text{gen.Proca}}^{\text{curved}} = -\frac{1}{4}F_{\mu\nu}^2 + \sum_{n=2}^{6} \beta_n\mathcal{L}_n
\end{align}

\noindent where the Lagrangians are organized in a derivative expansion as

\begin{align}
\mathcal{L}_2 &= G_2(A_\mu, F_{\mu\nu}, \tilde{F}_{\mu\nu}) \\
\mathcal{L}_3 &= G_3(X)\nabla_\mu A^\mu \\
\mathcal{L}_4 &= G_4(X)\bar{R} + G_{4,X}\left[(\nabla_\mu A^\mu)^2 - \nabla_\rho A_\sigma \nabla^\sigma A^\rho\right] \\
\mathcal{L}_5 &= G_5(X)\bar{G}_{\mu\nu}\nabla^\mu A^\nu - \frac{1}{6}G_{5,X}\left[(\nabla \cdot A)^3 \right. \\
&\quad \left. + 2\nabla_\rho A_\sigma \nabla^\gamma A^\rho\nabla^\sigma A_\gamma - 3(\nabla \cdot A)\nabla_\rho A_\sigma \nabla^\sigma A^\rho\right] \nonumber \\
&\quad - g_5(X)\tilde{F}^{\rho\mu}\tilde{F}_\mu^{\;\;\sigma}\nabla_\rho A_\sigma \nonumber \\
\mathcal{L}_6 &= G_6(X)\mathcal{L}^{\mu\nu\rho\sigma}\nabla_\mu A_\nu \nabla_\rho A_\sigma + \frac{G_{6,X}}{2}\tilde{F}^{\rho\sigma}\tilde{F}^{\mu\nu}\nabla_\rho A_\mu \nabla_\sigma A_\nu
\end{align}
where $G_i$ are arbitrary functions of $X=A_\mu A^\mu$, $G_{i,X}$ represents their derivatives with respect to X, and $\bar{G}_{\mu\nu}$ is the usual Einstein tensor and $\mathcal{L}^{\mu\nu\rho\sigma}=\frac{1}{4} \varepsilon^{\mu\nu\alpha\beta} \varepsilon^{\rho\sigma\gamma\delta}\bar{R}_{\alpha\beta\gamma\delta}$. 

Performing the Stückelberg trick to decompose $A_\mu\rightarrow A_\mu +\partial_\mu\varphi$, we can restore gauge invariance, in much the same way as we did in previous sections, before fixing unitary gauge. Here, the scalar $\varphi$ represents the longitudinal mode, and the gauge invariance allows us to keep only two propagating modes for $A_\mu$, thus preserving the three original modes of the original massive vector. Of course this is nothing but the old story of gauge symmetry being gauge redundancy. An interesting result is that in \cite{Heisenberg:2014rta}, it is shown that if we perform the Stückelberg trick, choosing $G_{2,3,4,5}(X)=X$, in the longitudinal limit where $A_\mu\rightarrow0$, Galileons emerge.

We now show a table categorizing each of the leading non-trivial interactions according to which Proca sub-family they fall under. The full index form of these terms can be found in Appendix \ref{Appexpansion}. Integration by parts might be necessary at times to cast them into this form though it is a straightforward computation.

\begin{table}[h]
\centering
\small
\setlength{\tabcolsep}{7pt}
\renewcommand{\arraystretch}{1.22}

\newcommand{\yes}{\(\checkmark\)}
\newcolumntype{Y}{>{\centering\arraybackslash}m{2.6em}}

\arrayrulecolor{black!30}
\newcommand{\softmidrule}{\arrayrulecolor{black!14}\midrule\arrayrulecolor{black!30}}

\begin{tabularx}{\textwidth}{@{}>{\raggedright\arraybackslash}X
  !{\vrule width 0.9pt} *{6}{Y}@{}}
\toprule
\textbf{Interaction}
  & \(\boldsymbol{G_2}\) & \(\boldsymbol{G_3}\)
  & \(\boldsymbol{G_4}\) & \(\boldsymbol{G_5}\)
  & \(\boldsymbol{g_5}\) & \(\boldsymbol{G_6}\) \\
\arrayrulecolor{black!45}
\Xhline{0.9pt}
\noalign{\vskip -1.2pt}
\Xhline{0.9pt}
\arrayrulecolor{black!30}

\(F \wedge \star F\)                                           & \yes &     &     &     &     &     \\
\softmidrule
\(A \wedge \star A\)                                          & \yes &     &     &     &     &     \\
\softmidrule
 \(R\equiv\varepsilon_{abcd}\, R^{ab} \wedge E^{c} \wedge E^{d}\)        & \yes & \yes & \yes &     &     &     \\
\softmidrule
\(R^{ab} \wedge A \wedge \star(A \wedge E_{a} \wedge E_{b})\) & \yes & \yes & \yes &     &     &     \\
\softmidrule
\(R_{ab} \wedge F \wedge \star(F \wedge E^{a} \wedge E^{b})\) & \yes &     &     &     & \yes & \yes \\
\softmidrule
\(\mathcal{L}_{GB}=\varepsilon_{abcd}\, R^{ab} \wedge R^{cd}\)                    &     &     & \yes & \yes &     &     \\
\softmidrule
\(P=R^{ab} \wedge R_{ab}\)                                      & \yes &     &     &     &     &     \\
\softmidrule
\(R^{ab} \wedge E_{b} \wedge \star(R_{cb} \wedge E^{c})\)     & \yes &     &     &     &     &     \\
\softmidrule
\(R^{ab} \wedge E_{a} \wedge \star(E_{b} \wedge F)\)          & \yes &     &     &     &     &     \\
\softmidrule
\(\varepsilon_{abcd}\, R^{ab} \wedge E^{c} \wedge \star(E^{d} \wedge F)\)
                                                              & \yes &     &     &     &     &     \\
\bottomrule
\end{tabularx}
\caption{Differential form terms across \(G_i\) sectors. See Appendix \ref{Appexpansion} for index notation of the different terms.}
\label{tab:diff-forms-Gi}
\end{table}

A perhaps surprising result is that we are yet to see terms in the $G_5(X)$ sector of the generalized Proca theory appear in the broken Weyl interactions. The only place for it to be found in the leading interactions of this Vector-Tensor theory is in the Gauss-Bonnet term, which is topological and a total derivative. 

In all cases checked thus far, all parity-preserving interactions
 adjust to the structure of generalized Proca theories, which is extremely restrictive. It would be interesting to have a proof of this feature for all higher dimensional operators in the EFT.

\section{Torsion}
\label{sec:Torsion}

Torsion, absent in General Relativity, enters naturally in Einstein–Cartan theory: Cartan’s \cite{Cartan1922} 1922 extension of the Levi-Civita connection introduced a non-vanishing antisymmetric part (torsion) and linked it to rotational momentum; the idea went largely unnoticed in the physics community until Kibble \cite{Kibble1961} derived the Einstein-Cartan equations by gauging the local Poincaré group and Sciama \cite{Sciama1964} showed that spin, alongside energy–momentum, sources the gravitational field. This was an important advance, as it allowed one to describe gravity in the same ubiquitous gauge-theoretic footing as the other fundamental interactions.

Here, we write a short aside on the role of torsion in this theory. We applied the covariant constraint
\begin{equation}
    T^a= \bar{T}^a+E^a\wedge A=0
\end{equation}
To see the implications of this single constraint in the GR torsion, we decompose this into three components;
\begin{align}
\bar{T}^{(1)\lambda}{}_{\mu\nu} &\;\equiv\;
      \tfrac12\,\bar{T}^{\lambda}{}_{\mu\nu}
    + \tfrac12\,\bar{T}^{\lambda}{}_{\nu\mu}
    + \tfrac16\,g_{\mu\nu}\,\bar{T}^{(2)\lambda}
    - \tfrac16\,\delta^{\lambda}{}_{\nu}\,\bar{T}^{(2)}_{\mu}
    + \tfrac13\,\delta^{\lambda}{}_{\mu}\,\bar{T}^{(2)}_{\nu}, \\[4pt]
\bar{T}^{(2)}_{\mu} &\;\equiv\;
      \bar{T}^{\lambda}{}_{\lambda\mu}, \\[4pt]
\bar{T}^{(3)}_{\mu} &\;\equiv\;
      \tfrac16\,\varepsilon_{\mu\nu\sigma\lambda}\,\bar{T}^{\nu\sigma\lambda}.
\end{align}
By setting  $T^a\equiv 0$, it is straightforward to show that this condition is equivalent to $\frac{1}{3}\bar{T}^{(2)}_{\mu}=
A_\mu$, and that $\bar{T}^{(1)\lambda}{}_{\mu\nu}=\bar{T}^{(3)}_{\mu} =0$ . This is interesting for two reasons: first, there has been shown that most torsionful theories, or theories non-linear in curvature often come accompanied by instabilities \cite{instabilityStelle:1977ry,torsionghostHECHT199012,R^2PhysRevD.21.3269}  and only some scalar modes have been shown to propagate without such pathologies \cite{acausalChen:1998ad,torsionokHecht:1996np}. In \cite{Barker:2023fem}, the authors show that incorporating a term $\alpha \bar{R}_{[\mu\nu]}\bar{R}^{[\mu\nu]}$, together with a $ M_{\mathrm{Pl}}{ }^{2(2)} \bar{T}_\mu{ }^{(2)} \bar{T}^\mu$ solves both the issue of ghosts, and that of strong coupling. In our Weyl broken gravity, we find what, as far as we know is, the only realization of this construction without the introduction of Lagrange multipliers. The term $ R_{[\mu\nu]}$ simplifies to $F_{\mu\nu}$ in this theory, and so $\alpha \bar{R}_{[\mu\nu]}\bar{R}^{[\mu\nu]}$ is included in our construction automatically as the Proca kinetic term. In addition, the coefficient $\alpha$ must be negative \cite{Barker:2023fem}, which is the right sign for a healthy Spin-1 field. It is remarkable that this theory falls out so very naturally from requiring $T^a=0$, satisfying all the aforementioned constraints for a healthy theory of propagating torsion. In our effective Lagrangian, we could always use $\bar{T}^a=-E^a\wedge A$ to trade all Weyl gauge fields $A$ for the torsion $\bar{T}^{(2)}_{\mu}$ and instead interpret this theory as a theory of propagating torsion.

\section{Approximate symmetries}
\label{sec:approximatesymmetries}

Approximate Weyl symmetry breaking can play a key role near RG fixed points in gravitational theories. In \cite{Iorio:1996ad}, it is shown that, provided a CCJ–improved stress energy tensor \cite{Callan:1970ze} exists, a necessary and sufficient condition for promoting a flat spacetime theory to a Weyl invariant theory is conformal invariance in flat space\footnote{Together with coupling to $G_{\mu\nu}$, reminiscent of the generalized Proca coupling in $\mathcal{L}_4$.}. From our previous discussion it follows that not all scale invariant theories can be promoted to Weyl invariance once curvature is turned on, although a sizeable subset can.

Many systems exhibit hard scale symmetry breaking. For example, classically scale invariant theories such as massless QCD generate a scale $\Lambda_{QCD}$ through dimensional transmutation. Along RG flows the breaking typically has coefficients of order one, and the trace anomaly is directly tied to the beta functions,

\begin{equation}
T^\mu{}_\mu \;=\; \sum_i \beta_i(g)\,\mathcal{O}^i \;+\; \text{(curvature anomaly)} \, ,
\end{equation}

where the curvature term depends on the spacetime dimension; in two dimensions one has $\tfrac{c_{\text{tot}}}{24\pi} R$. In theories with special matter content, such as the bosonic string or superstring, the curvature anomaly cancels, and the quantum trace vanishes only when the beta functions vanish. There is no spurion that can be introduced to remove this effect, which marks a genuinely hard departure from scale invariance when $\beta_i \neq 0$.

Nevertheless, near an RG fixed point $g=g^*$ one may linearize the flow,

\begin{equation}
    \beta_i \;=\; B_{ij}\,(g-g^*)^j \, ,
\end{equation}

with $B_{ij}$ a matrix of coefficients. This linearization suppresses the otherwise hard breaking and yields an approximate Weyl symmetry in the relevant regime. Motivated by this, a natural direction is to identify theories that display large hierarchies of scales sourced by small breakings of Weyl symmetry under the conditions outlined above. This perspective suggests a pathway to explaining hierarchies in a broad class of models.

As a short remark and that may help motivate future work in this direction we consider the Generalized Proca case.
As shown previously, the mixed–curvature term
\begin{equation}
 R^{ab}\!\wedge F\!\wedge\!\star\!\bigl(F\!\wedge E_a\!\wedge E_b\bigr)
\end{equation}
reproduces the $g_5(X)$ piece of $\mathcal{L}_5$.  
We are yet to find a contraction that generates terms of the $G_5(X)$ branch using the building blocks employed so far, that also does not lead to a ghost. We can find such terms in Lagrangians with Kretschmann-like terms, but as aforementioned, these inevitably lead to ghosts on the gravity sector, and so we choose not to include them in a sensible construction. Finally, the Gauss-Bonnet terms do include these contributions, but are total derivative topological terms.

The relationship between Galileon and Weyl symmetries provides insight into potential hierarchies in Generalized Proca theories. The Stückelberg replacement $A_\mu \to A_\mu + \partial_\mu\pi$ introduces a longitudinal mode with Galileon symmetry $\pi \to \pi + c + b_\mu x^\mu$. This transformation can be seen as a truncation at linear order in spacetime coordinates of the Weyl transformation of the dilaton  $\lambda(x) = c + b_\mu x^\mu$ in the dilaton shift $\varphi \to \varphi + \Lambda(x)$ under metric rescaling $g_{\mu\nu} \to e^{2\Lambda(x)}g_{\mu\nu}$.

This symmetry hierarchy suggests an intriguing possibility: the conspicuous absence of $G_5(X)$ terms in our broken Weyl construction may not be accidental. If including $G_5(X)$ breaks the enhancement from Galileon to Weyl symmetry, then 't Hooft naturalness arguments~\cite{tHooft:1979rat} would predict a hierarchy between Wilson coefficients. Setting $G_5(X) = 0$ would enlarge the symmetry from Galileon to Weyl, making small values of $G_5(X)$ technically natural as they would be enlarged by radiative corrections through renormalization. In contrast, $g_5(X)$, and other terms in Generalized Proca appear readily and enjoy no such symmetry protection.
The fact that \(G_5(X)\) appears in our construction only through the Gauss-Bonnet density in \(d=4
\) (a boundary term) or through curvature squared couplings that propagate the massive spin 2 ghost supports a symmetry based suppression of \(G_5\). In the limit \(G_5 \to 0\) the theory exhibits an enhanced approximate Weyl symmetry via the St\"uckelberg longitudinal mode, which makes a small \(G_5\) technically natural in the sense of 't Hooft. A useful next step would be a mathematical proof that this symmetry indeed forbids the \(G_5\) operator at the relevant orders in the EFT. Finally, for \(d>4\) the Gauss-Bonnet term is neither topological nor a total derivative and leads to second order equations, so the associated \(G_5\) like structure becomes a bona fide interaction in higher dimensions.

A different extension would follow \cite{Goon:2014ika,Goon:2014paa}: viewing a local gauge group as an infinite tower of global symmetries recovers non-Abelian. Yang-Mills from \( SU(N)_{\text{local}}\!\to\!SU(N)_{\text{global}} \), and in the Abelian case reproduces the usual Stückelberg description. Related work shows how Einstein gravity can be obtained as a nonlinear realization of diffeomorphisms. An analogous cascade,
\[
  \text{Weyl}_{\text{local}}\ \longrightarrow\ \text{Galileon}_{\text{global}}
\]
remains open. Since the ordinary Galileon keeps the constant and linear pieces of \( \Lambda(x) \) while the special Galileon retains also the quadratic piece in \( x \), one can imagine a  breaking with progressively fewer broken modes. Hierarchies among Galileon operators then connect naturally to the weakly broken Galileon framework, where approximate symmetry enforces controlled, radiatively stable deviations \cite{Pirtskhalava:2015nla}.

\section{Discussion}
\label{sec:discussion}

 In this work, we have motivated the theoretical appeal of gauge-gravity theories featuring non-linearly realized Weyl symmetry. We have demonstrated how, in this framework, the Weyl gauge field is Higgsed, with the dilaton embodied naturally as a Stückelberg field. After being "eaten" by the gauge field, the dilaton acts as the longitudinal mode of $A_\mu$, exhibiting Galileon-like properties. The massive vector acquires a mass of order $\mathcal{O}(M_{\text{Pl}})$, barring any fine-tuning of the universal coupling $q$ of the Weyl gauge field to other fields. We remark that the physical mass of the Proca field will generally be below the Planck mass; in a particle physics context, the gauge field is rescaled, $A_\mu\rightarrow q A_\mu$, with $q$ the coupling of the Proca field. Therefore, after the appropriate change of variables, the mass of the Proca field will be given by $m\sim qM_{\rm Pl}$. This is directly analogous to fermion masses being proportional to the Higgs VEV times their Yukawa coupling.

This theory possesses a rich structure, allowing us to build bridges between different "continents" of modified gravity. 
Our first bridge links to generalized Proca theories. We have shown that the leading 
interactions 
fall precisely within the severely constrained family of generalized Proca interactions which are ghost-free.
This is a highly non-trivial result given the finely tuned interactions that generalized Proca requires. Secondly, we have shown that upon imposing the covariant constraint $T^a=0$, one can naturally identify the vector component of general relativity torsion with the Weyl gauge field,  eliminating tensor and pseudo-vector components, all in one swift stroke. Previous literature \cite{Barker:2023fem} identified the vector component in this decomposition as the unique ghost-free propagation of vector torsion. Given the properties of generalized Proca, it further solidifies the claim that the propagating torsion theory presented here is devoid of ghost instabilities.

These vector bosons, although still very massive given the large hierarchy between the Planck mass and the electroweak scale, are still parametrically smaller than the quantum gravity scale $M_{\rm Pl}$ and couplings $q \sim\mathcal{O}(10^{-2})$ would still be technically natural and have observable effects within the EFT. No screening mechanism such as Vainshtein \cite{Vainshtein:1972sx} is necessary for such theories as the Compton radius $r_{\rm C}\sim \frac{1}{M_A}\sim \ell_{\rm Pl}$ is far smaller than any macroscopic length scale and so any forces mediated by the Weyl vector would be exponentially suppressed barring an extreme fine tuning of the aforementioned coupling $q$.

Massive Weyl fields have long been proposed as potential candidates for dark energy or dark matter. Given their high mass, it would be interesting to see if they can still be viable dark sector candidates.   At extremely high energies, these massive particles could have been efficiently produced if a phase transition occurred, leaving vestigial imprints of the field dynamics detectable through inflationary signatures. As a low-energy effective theory, the cutoff scale is $\Lambda \sim  q M_{\rm Pl} $ once the Proca field is integrated out. Higher order corrections will still be suppressed by energies that are far beyond what high energy astrophysical processes can achieve.  This is in alignment with the constraints on higher order curvature corrections suppressed by the cutoff using binary black hole mergers \cite{Sennett:2019bpc}. A gravitational wave background study of a phase transition remains to be done. Some phenomenological aspects of this particular theory have been studied for inflation \cite{Ghilencea:2018thl}.

This rich set of phenomenological applications and theoretical features makes non-linearly realized Weyl theories subject worthy of further research and development. As emphasized throughout the text, it would be particularly interesting to systematically analyze theories in which Weyl symmetry is only mildly broken. This could resolve whether the missing term $G_5(X)$ in Proca theories might finally emerge, potentially enabling significant hierarchies among these terms. Such big hierarchies of scale would be technically natural, protected by the approximate Weyl symmetry. On the other hand, terms such as $R^2$ propagate an extra scalar degree of freedom, and it would be interesting to see whether these do indeed fall under the beyond Generalized Proca \cite{Heisenberg:2016eld} theories including scalar modes, but which naively seem to have $G_5$-like interactions. 

While we have focused on a gauged dilaton theory, a similar treatment of other modifications of gravity, such as full conformal gravity, would also be of theoretical interest. In this particular case, the role of the dilaton might be obscured within the conformal building blocks due to an additional constraint that did not appear here, known as inverse Higgs constraint, but which is present in the conformal case given the commutator of generators of translations and special conformal transformations. Additionally, Wess-Zumino terms, absent in our current broken Weyl constructions, do in fact exist in theories with broken conformal symmetry \cite{Goon2012GalileonsTerms}, which we will tackle in future works.

\acknowledgments
We are thankful to Riccardo Penco, Ira Rothstein, Rachel Rosen and Lavinia Heisenberg for insightful discussions. This work is supported by the US Department of Energy grant DE-SC0010118.  LF is partially supported by the Rafael del Pino Foundation.

\appendix

\appendix

\section{Conventions}
\label{app:conventions}
\textit{Notation:}  Quantities with a bar denote the GR solution, such as the Ricci scalar $\bar R$ written solely in terms of the vierbein, while the one containing the Proca field is $R ( E_{\mu}^{\;a}, A_{\mu} )$. Antisymmetrization is denoted by square brackets, $M_{[\mu\nu]} \equiv \tfrac12\!\left( M_{\mu\nu} - M_{\nu\mu} \right),$ with the usual extension to higher-rank tensors, symmetrization is defined analogously.

We differentiate between space-time and local Lorentz indices as in \cite{Delacretaz:2014oxa}:

\begin{itemize}
	\item $\mu, \nu, \sigma, \rho ...$ denote space-time indices.
	\item $a, b, c, d ...$ denote Lorentz indices.
	\item $i, j, k, l ...$ denote spatial components of the Lorentz indices.
\end{itemize}

We denote the time of occurrence and the location in space of an event with the four component vector, $x^a = (x^0,x^1,x^2,x^3) = (t,\vec{x})$, and define the flat space-time interval, $\mathrm{d}s$, between two events, $x^a$ and $x^a+\mathrm{d}x^a$, by the relation

\begin{equation}
\mathrm{d}s^2 = -  \mathrm{d}t^2 + \mathrm{d} x^2 + \mathrm{d} y^2 + \mathrm{d} z^2 ,
\label{eq:space-timeinterval}
\end{equation}

\noindent which we write using the notation

\begin{equation}
\mathrm{d} s^2 = \eta_{ab}\mathrm{d}x^a\mathrm{d}x^b \,\mathrm{;} \,\,\,\,\,\;\;\;\;  \eta_{ab} = \mathrm{diag} (-1, +1, +1, + 1) .
\label{eq:lineinterv}
\end{equation}

\noindent A one form and its field strength are written as
\begin{align}
A &= A_\mu\, dx^\mu,\\
F &= dA
   = \partial_\mu A_\nu\, dx^\mu \wedge dx^\nu
   = \frac{1}{2}\big(\partial_\mu A_\nu - \partial_\nu A_\mu\big)\, dx^\mu \wedge dx^\nu
   = \frac{1}{2}\,F_{\mu\nu}\, dx^\mu \wedge dx^\nu,
\end{align}
with
\begin{equation}
F_{\mu\nu} \equiv \partial_\mu A_\nu - \partial_\nu A_\mu = 2\,\partial_{[\mu}A_{\nu]}.
\end{equation}

For an $n$ form
\begin{equation}
\omega = \frac{1}{n!}\,\omega_{i_1\cdots i_n}\,dx^{i_1}\wedge\cdots\wedge dx^{i_n},
\end{equation}
the exterior derivative $d:\Omega^n\to\Omega^{n+1}$ acts as
\begin{align}
d\omega
&= \frac{1}{n!}\,\partial_\mu \omega_{i_1\cdots i_n}\,
   dx^\mu \wedge dx^{i_1}\wedge\cdots\wedge dx^{i_n},
\end{align}
or in components
\begin{equation}
(d\omega)_{\mu i_1\cdots i_n} = (n+1)\,\partial_{[\mu}\omega_{i_1\cdots i_n]}.
\end{equation}
Similarly, the Hodge star operator acts as 

\begin{align}
\star: \quad \Omega^r(\mathcal{M}) & \rightarrow \quad \Omega^{m-r}(\mathcal{M}) \\
d x^{\mu_1} \wedge \ldots \wedge d x^{\mu_r} \rightarrow & \star\left(d x^{\mu_1} \wedge \ldots \wedge d x^{\mu_r}\right) \\
& \equiv \frac{1}{(m-r)!} \varepsilon^{\mu_1 \ldots \mu_r}{ }_{\mu_{r+1} \ldots \mu_m} d x^{\mu_{r+1}} \wedge \ldots \wedge d x^{\mu_m}
\end{align}
An example of this would be 
\begin{equation}
    \tilde{F}=\star F=\frac{1}{4}\varepsilon_{\alpha\beta\mu\nu}F^{\mu\nu}dx^\alpha\wedge dx^\beta
\end{equation}
which in components yields $\tilde{F}^{\alpha\beta}=\frac12\varepsilon_{\alpha\beta\mu\nu}F^{\mu\nu}$ as usual.

\section{Generators and algebra}
\label{app:Commutators}
 Here we present all the operators we use and their commutators for future reference. 
The representation of the generators of the Lorentz and dilation group we use are 
\begin{align}
     & P_\mu=-i \partial_\mu  \\ 
     & D=-i x^\mu \partial_\mu \\ 
     & J_{\mu \nu}=i\left(x_\mu \partial_\nu-x_\nu \partial_\mu\right)
\end{align}
It is straightforward to see that $[\partial_\mu,x^\nu]=\delta_\mu^{\;\;\nu}$. With this we can derive 

\begin{align} 
{\left[D, P_\mu\right] } & =i P_\mu  \\
{\left[P_\rho, J_{\mu \nu}\right] } & =i\left(\eta_{\rho \mu} P_\nu-\eta_{\rho \nu} P_\mu\right) \\
{\left[J_{\mu \nu}, J_{\rho \sigma}\right] } & =i\left(\eta_{\nu \rho} J_{\mu \sigma}+\eta_{\mu \sigma} J_{\nu \rho}-\eta_{\mu \rho} J_{\nu \sigma}-\eta_{\nu \sigma} J_{\mu \rho}\right)
\end{align}

\section{Chevalley-Eilenberg cohomology}\label{CE cohomology}

As previously explained, the coset construction allows us to build invariant Lagrangians under the symmetries of our theory. However, this method inherently misses certain essential contributions: those terms whose variation under a symmetry transformation is a total derivative. 

Crucially, the $d$-form $\alpha$ itself is not necessarily invariant under all symmetries $G$. Instead, its variation under a transformation by an element of $G$, denoted $\delta_G \alpha$, is of exact form (i.e., $\delta_G \alpha = d\xi$ for some $(d-1)$-form $\xi$). This property ensures that the action $\int_{\mathcal{N}} \alpha$ remains invariant under $G$ (up to boundary terms, if $\partial\mathcal{N} \neq \emptyset$). The coset construction, which yields strictly invariant forms, would miss these WZ terms. The systematic characterization of such terms was provided by D'Hoker and Weinberg, and utilized by Witten \cite{DHoker:1994rdl,Witten:1983tw}, and later used by Goon \emph{et al.} in \cite{Goon2012GalileonsTerms} in the context of Galileons. Two of the seminal papers of Lie algebra and relative Lie algebra cohomology in the high energy physics literature are \cite{deAzcarraga:1998uy,deAzcarraga:1997qrt}. For this venture, the machinery of relative Lie-algebra (ChE) cohomology \cite{Chevalley:1948zz} plays a main role, which we briefly discuss below. 

\subsection*{Lie Algebra cohomology}
The following two sections summarize and build upon the presentation of Lie algebra cohomology in \cite{deAzcarraga:1998uy,Goon2012GalileonsTerms}.
Let \(\mathfrak g\) be the Lie algebra of a Lie group \(G\). Cochains can be defined over vector modules that transform in different representations, but given the fact that the objects we are trying to describe are Lagrangian densities, which are singlets under the action of \(G\), we will take the trivial representation. This amounts to defining our cochains with coefficients in \(\mathbb{R}\). This simplifies matters as the coboundary operator will take a familiar form. Then, for us, an \(n\) cochain in the trivial representation is defined as an alternating multilinear map \(\omega:\ \bigwedge^{n}\mathfrak g \to \mathbb{R}\), and the space of \(n\) cochains is \(C^n(\mathfrak g,\mathbb{R})=\Lambda^n \mathfrak g^{*}\). There exists a one to one correspondence between \(n\) cochains and left invariant (LI) \(n-\) forms. The coboundary operator \(\delta:C^{n}(\mathfrak g,\mathbb{R})\to C^{n+1}(\mathfrak g,\mathbb{R})\) also corresponds to the usual de Rham exterior derivative on LI forms d. In the trivial representation, the coboundary operator acts as
\begin{equation}
\delta \omega\left(X_1, X_2, \ldots, X_{n+1}\right)=\sum_{\substack{j, k=1 \\ j<k}}^{n+1}(-1)^{j+k} \omega\left(\left[X_j, X_k\right], X_1, \ldots, \hat{X}_j, \ldots, \hat{X}_k, \ldots, X_{n+1}\right)
\end{equation}
where $[\cdot,\cdot]$ is the Lie bracket, or in its matrix formulation, the commutator of Lie algebra elements. 
When working with an explicit basis $\{E_i\}$ for $\mathfrak{g}$ satisfying $[E_i,E_j]=c_{ij}^{\;\;k}E_k$ with $ c_{ij}^{\;\;k}$ being the usual structure constants, we can represent the action of the coboundary operator on cochains in a particularly nice way.
For trivial coefficients, the coboundary differential $\delta$ acts on a $1$-cochain $\omega$ by
\begin{equation}
  (\delta\omega)(X,Y)=-\,\omega([X,Y]) .
  \label{coboundary action}
\end{equation}
we can then evaluate $\delta\omega^i$ on basis elements:
\begin{align}
  (\delta\omega^i)(E_j,E_k)
  &= -\,\omega^i([E_j,E_k]) = -\,\omega^i\!\left(c^{\,m}{}_{jk}\,E_m\right)
   = -\,c^{\,i}{}_{jk}.
\end{align}
Writing $\delta\omega^i$ in the basis of $2$ forms
\begin{equation}
  \delta\omega^i=\frac12\,A^{\,i}{}_{jk}\,\omega^j\wedge\omega^k .
\end{equation}
and comparing with $(\delta\omega^i)(E_j,E_k)=-\,c^{\,i}{}_{jk}$ yields
$A^{\,i}{}_{[jk]}=-\,c^{\,i}{}_{jk}$. This can be easily extended to the action on several forms. Finally, it is easy to show that the coboundary operator $\delta$ is nilpotent, i.e. $\delta^2=0$ using the Jacobi identity. Equipped with the coboundary $\delta$, the $n$th cohomology measures cocycles ($\delta \omega=0$) modulo coboundaries ($\omega=\delta \alpha, \omega\in \Lambda^n \mathfrak g^{*}, \alpha \in \Lambda^{n-1} \mathfrak g^{*}$):
\begin{equation}
H^n(\mathfrak g;\mathbb R)
=\frac{\ker\bigl(\delta:\Lambda^n\mathfrak g^{*}\to\Lambda^{n+1}\mathfrak g^{*}\bigr)}
       {\operatorname{im}\bigl(\delta:\Lambda^{n-1}\mathfrak g^{*}\to\Lambda^n\mathfrak g^{*}\bigr)}.
\end{equation}
Equivalently, two cocycles $\omega,\omega'\in\Lambda^n\mathfrak g^{*}$ are identified if they differ by a coboundary:
$\omega'\sim\omega$ when $\omega'-\omega=\delta\alpha$ for some $\alpha\in\Lambda^{n-1}\mathfrak g^{*}$.

\subsection*{Relative Lie algebra cohomology}
One refinement that we can make to Lie algebra cohomology is to define it for the pair $(\mathfrak g,\mathfrak h)$ associated with the homogeneous space $G/H$.
For Lie algebras $\mathfrak{h}\subset\mathfrak{g}$, the space of relative $n$-cochains $\Lambda^n(\mathfrak{g},\mathfrak{h})$ consists of alternating multilinear maps from $\bigwedge^{n} \mathfrak{g}$ to $\mathbb{R}$ that satisfy two conditions, which we will motivate below.

Heuristically, to descend to the quotient space $\mathfrak{g}/\mathfrak{h}$, elements are equivalence classes $[X] = X + \mathfrak{h}$, where $X_1 \sim X_2$ if $X_1 - X_2 \in \mathfrak{h}$. For a map to be insensitive to the choice of representative (and hence descend), it should give the same result after replacing any argument by an $\mathfrak h$-shift.

If we replace an argument $X_1$ with an equivalent element $X_1' = X_1 + V$ with $V \in \mathfrak{h}$, then by multilinearity:
\begin{equation}
         \omega_n(X_1 + V, X_2, \ldots, X_n) = \omega_n(X_1, X_2, \ldots, X_n) + \omega_n(V, X_2, \ldots, X_n).
\end{equation}

For the form to be independent of the representative, we need
\begin{equation}
    \omega_n(X_1 + V, X_2, \ldots, X_n) = \omega_n(X_1, X_2, \ldots, X_n),
\end{equation}
for $\omega_n \in \Lambda^n\mathfrak g$ which requires $\omega_n(V, X_2, \ldots, X_n) = 0$ for all $V \in \mathfrak{h}$ and any $X_i \in \mathfrak{g}$. This is the \emph{vanishing} (horizontality) condition. There is a one-to-one correspondence between $n$-cochains and left-invariant $n$-forms on $G$; the condition above simply states that the interior product, or contraction of our forms with vectors along the $\mathfrak h$ directions vanishes.

Secondly, the \textit{invariance condition} ensures that our forms are preserved under the natural action of $\mathfrak{h}$ on $\mathfrak{g}/\mathfrak{h}$. The adjoint action of $\mathfrak{h}$ on $\mathfrak{g}$ is given by $\text{ad}_V(X) = [V,X]$ for $V \in \mathfrak{h}$. This action preserves the quotient structure: if $X_1 \equiv X_2 \pmod{\mathfrak{h}}$, then
\begin{equation}
    [V, X_1] - [V, X_2] = [V, X_1 - X_2] \in \mathfrak{h},
\end{equation}
since $\mathfrak{h}$ is a subalgebra and $X_1 - X_2 \in \mathfrak{h}$. For a form to be invariant under this action, its Lie derivative with respect to any $V \in \mathfrak{h}$ must vanish:
\begin{equation}
    \sum_{i=1}^n \omega_n(X_1, \ldots, [V, X_i], \ldots, X_n) = 0,
\end{equation}
which ensures that the form transforms trivially under the action of $\mathfrak{h}$, making it a well-defined object associated with the quotient space $\mathfrak{g}/\mathfrak{h}$. 
To construct a cohomology theory from these relative cochains, we need a differential operator that maps $n$-cochains to $(n+1)$-cochains while preserving the defining conditions. This is provided by the Chevalley-Eilenberg coboundary operator
\begin{equation}
    \delta\colon \Lambda^n(\mathfrak{g},\mathfrak{h}) \longrightarrow \Lambda^{n+1}(\mathfrak{g},\mathfrak{h}),
\end{equation}
which is the analogue of the exterior derivative in de Rham cohomology. Crucially, this operator automatically preserves horizontality and, when restricted to relative cochains, maps invariant cochains to invariant ones.
 The relative cohomology is then defined as
\begin{equation}
    H^n(\mathfrak{g}, \mathfrak{h})=\frac{\ker\delta_n\left(\Lambda^n(\mathfrak{g}, \mathfrak{h})\right)}{\text{Im}\,\delta_{n-1}\left(\Lambda^{n-1}(\mathfrak{g}, \mathfrak{h})\right)}.
\end{equation}

The non-trivial elements of $H^{d+1}(\mathfrak{g},\mathfrak{h})$ provide an \emph{upper bound} on the number of WZ terms in $d$ dimensions \cite{Goon2012GalileonsTerms,deAzcarraga:1998uy,deAzcarraga:1997qrt}. Several mechanisms can cause cohomologically distinct forms to yield identical or trivial physical contributions. Inverse Higgs constraints impose algebraic relations on the Maurer-Cartan form, so forms that are cohomologically non-trivial before applying these constraints may become exact (total derivatives) afterward, contributing nothing to the action \cite{Goon2012GalileonsTerms,deAzcarraga:1998uy}. Finally, multiple cohomologically distinct forms may become equivalent after applying standard field-theory manipulations such as integration by parts and field redefinitions.

In fibre-bundle language, we view $G$ as a principal $H$-bundle over the base space $G/H$, where $G/H$ is a homogeneous space, in other words, locally we have a trivial $G/H \times H$ structure. To see how CE cohomology maps to differential forms, we introduce basic forms.  A differential form on $G$ is basic if it satisfies two key properties:
(i) it is horizontal, meaning it vanishes when any of its arguments is a vector field tangent to the $H$-fibres (i.e., $i_V\omega = 0$ for any vector $V$ generating the $H$-action along the fibres, corresponding to elements in $\mathfrak{h}$). This is clearly the vanishing condition. 
(ii) It is invariant under the right $H$-action on $G$. This physically tells us that we can move the differential form around the fiber (act on it with the H subgroup), and the form will be unchanged. This is telling us that $\mathcal{L}_V\omega=0$. 

So we see that the two conditions outlined above. Relative Lie algebra $k$-cochains $\Omega^k(\mathfrak{g},\mathfrak{h})$ can be identified with left $G$-invariant differential $k$-forms on the group manifold $G$ that are also \emph{basic} with respect to the $H$-fibration. 
Such left $G$-invariant basic forms on $G$ descend uniquely to $G$-invariant differential forms on the coset space $G/H$. The coboundary operator $\delta$ on cochains then corresponds to the exterior derivative $d$ on these forms.

In practical terms, to compute the relative cohomology $H^{d+1}(\mathfrak g,\mathfrak h)$ we can apply this recipe:
\begin{enumerate}
  \item  Work out the action of the CE differential on the dual basis one–forms (see Eq.~\eqref{coboundary action}): $\delta \omega^{A}=-\tfrac12 f^{A}{}_{BC}\,\omega^{B}\wedge\omega^{C}$.
  \item Write all candidates in $\Lambda^{d+1}\mathfrak g^{*}$ built only from coset one–forms (no $\mathfrak h$ legs). Equivalently impose $i_{V}\omega=0$ and $\mathcal L_{V}\omega=0$ for all $V\in\mathfrak h$. Keep only those with $H$–invariant coefficient tensors.
  \item Apply $\delta$ to each candidate and project to the relative complex by discarding any term that contains a single $\mathfrak h$ one–form. Survivors form $Z^{d+1}(\mathfrak g,\mathfrak h)$.
  \item
  List all relative $d$–cochains in $\Lambda^{d}\mathfrak g^{*}$ apply $\delta$. After projecting out the forms with $\mathfrak h$ legs, these span $B^{d+1}(\mathfrak g,\mathfrak h)=\delta\big(\Lambda^{d}\mathfrak g^{*}\big)_{\text{rel}}$.
  \item The relative cohomology is $H^{d+1}(\mathfrak g,\mathfrak h)=Z^{d+1}(\mathfrak g,\mathfrak h)\big/ B^{d+1}(\mathfrak g,\mathfrak h)$; the candidates from step 2 that are cocycles in step 3 but not coboundaries in step 4.
\end{enumerate}
\newpage
\section{Differential form to index notation dictionary}\label{Appexpansion}

\begin{table}[H]
\centering
\small
\setlength{\tabcolsep}{7pt}
\renewcommand{\arraystretch}{1.18}

\newcolumntype{L}{>{\raggedright\arraybackslash}X}
\arrayrulecolor{black!30}
\newcommand{\softmidrule}{\arrayrulecolor{black!14}\midrule\arrayrulecolor{black!30}}

\begin{tabularx}{\textwidth}{@{}L !{\vrule width 0.9pt} L@{}}
\toprule
\textbf{Differential form} & \textbf{Index form} \\
\arrayrulecolor{black!45}\Xhline{0.9pt}\noalign{\vskip -1.2pt}\Xhline{0.9pt}\arrayrulecolor{black!30}

\(\mathcal{L}_{\mathrm{EH}}=\varepsilon_{abcd}\,R^{ab}\wedge E^{c}\wedge E^{d}\)
&
\(\displaystyle E\bigl(\bar R + 6\,\bar\nabla_\mu A^\mu - 6\,A_\mu A^\mu\bigr)\)
\\ \softmidrule

\(N \equiv R^{ab}\wedge E_a \wedge E_b\)
&
\(\displaystyle \varepsilon_{\gamma\mu\xi\sigma}\,R^{\gamma\mu\xi\sigma}=0\)
\\ \softmidrule

\(P \equiv R^{ab}\wedge R_{ab}\)
&
\(\displaystyle \tfrac18\,\widetilde{\bar R}_{\mu\nu\alpha\beta}\,\bar R^{\mu\nu\alpha\beta}
+ \tfrac14\,\tilde F^{\mu\nu} F_{\mu\nu}\)
\\ \softmidrule

\(\mathcal{E}\equiv \varepsilon_{abcd}\,R^{ab}\wedge R^{cd}\)
&
\(\displaystyle
E\Bigl(
\bar R^{2}
- 4\,\bar R^{\mu\nu}\bar R_{\mu\nu}
+ \bar R_{\mu\nu\alpha\beta}\bar R^{\mu\nu\alpha\beta}
- 8\,\bar G_{\mu\nu} A^\mu A^\nu
- 4\,A^{2}\bar R
- 8\,\bar G_{\mu\nu}\,\bar\nabla^{\mu} A^{\nu}
\Bigr)\)
\\ \softmidrule

\(K \equiv R^{ab}\wedge E^{c}\wedge\star(R_{ab}\wedge E_{c})\)
&
\(\displaystyle
E\Bigl(
12\,A^{4}
+ 8\,A^{\mu}A^{\nu}\bar G_{\mu\nu}
+ \bar R_{\alpha\beta\mu\nu}\bar R^{\alpha\beta\mu\nu}
- 24\,A^{2}\bar\nabla_{\gamma}A^{\gamma}
+ 4\,\bar\nabla_{\alpha}A^{\alpha}\,\bar\nabla_{\gamma}A^{\gamma}
+ 8\,\bar R_{\alpha\gamma}\,\bar\nabla^{\gamma}A^{\alpha}
+ 8\,\bar\nabla_{\gamma}A_{\alpha}\,\bar\nabla^{\gamma}A^{\alpha}
\Bigr)\)
\\ \softmidrule

\(R^{ab}\wedge E_{b}\wedge\star(R_{cb}\wedge E^{c})\)
&
\(\displaystyle E\,F^{\mu\nu}F_{\mu\nu}\)
\\ \softmidrule

\(R^{ab}\wedge E^{c}\wedge\star(R_{ac}\wedge E_{b})\)
&
\(\displaystyle
E\Bigl(
- 6\,A^{4}
- \bar R_{\alpha\gamma}\bar R^{\alpha\gamma}
+ 2\,A^{2}\bar R
+ \tfrac12\,\bar R_{\alpha\beta\mu\nu}\bar R^{\alpha\beta\mu\nu}
- 2\,\bar R\,\bar\nabla_{\alpha}A^{\alpha}
+ 12\,A^{2}\bar\nabla_{\gamma}A^{\gamma}
- 6\,\bar\nabla_{\alpha}A^{\alpha}\,\bar\nabla_{\gamma}A^{\gamma}
\Bigr)\)
\\ \softmidrule

\(\mathcal{L}_{\mathrm{EH}}\wedge \star \mathcal{L}_{\mathrm{EH}}\)
&
\(\displaystyle
E\Bigl(
\bar R^{2}
+ 36\,A^{4}
- 12\,A^{2}\bar R
+ 12\,\bar R\,\bar\nabla_{\alpha}A^{\alpha}
- 72\,A^{2}\bar\nabla_{\gamma}A^{\gamma}
+ 36\,(\bar\nabla_{\alpha}A^{\alpha})^{2}
\Bigr)\)
\\ \softmidrule

\(R^{ab}\wedge E_{a}\wedge\star(E_{b}\wedge F)\)
&
\(\displaystyle E\,F^{\mu\nu}F_{\mu\nu}\)
\\ \softmidrule

\(\varepsilon_{abcd}\,R^{ab}\wedge E^{c}\wedge\star(E^{d}\wedge F)\)
&
\(\displaystyle \propto\ \tilde F_{\mu\nu}\,F^{\mu\nu}\)
\\ \softmidrule

\(\varepsilon_{abcd}\,R^{ab}\wedge A\wedge\star(E^{c}\wedge E^{d}\wedge A)\)
&
\(\displaystyle 0\)
\\ \softmidrule

\(R^{ab}\wedge A\wedge\star(A\wedge E_{a}\wedge E_{b})\)
&
\(\displaystyle
E\Bigl(
3\,A^{4}
+ A^{\mu}A^{\nu}\bar G_{\mu\nu}
+ 2\,A^{\mu}A^{\nu}\bar\nabla_{\nu}A_{\mu}
- 2\,A^{2}\bar\nabla_{\nu}A^{\nu}
\Bigr)\)
\\ \softmidrule

\(R_{ab}\wedge F\wedge\star(F\wedge E^{a}\wedge E^{b})\)
&
\(\displaystyle
E\Bigl(
\mathcal{L}^{\alpha\beta\mu\nu}\,F_{\alpha\beta}F_{\mu\nu}
+ A^{\alpha}A^{\beta}F_{\alpha}{}^{\mu}F_{\beta\mu}
- \tfrac12\,A^{2}F_{\beta\mu}F^{\beta\mu}
+ F_{\alpha}{}^{\mu}F_{\beta\mu}\,\bar\nabla^{\beta}A^{\alpha}
\Bigr)\)
\\ \softmidrule

\(F\wedge A\wedge E_{a}\wedge\star(R^{ab}\wedge E_{b}\wedge A)\)
&
\(\displaystyle
E\Bigl(
A^{\alpha}A^{\beta}F_{\beta\mu}F^{\mu}{}_{\alpha}
+ \tfrac12\,A^{2}F_{\mu\nu}F^{\mu\nu}
\Bigr)\)
\\
\bottomrule
\end{tabularx}

\caption{Side-by-side mapping from differential-form expressions to their index forms up to an overall multiplicative factor. In the second line we have used the first algebraic Bianchi identity \(R_{\mu[\nu\alpha\beta]}=0\) for instructive purposes, and it is also used elsewhere.}
\label{tab:forms-vs-index}
\end{table}

\bibliographystyle{JHEP}
\bibliography{bib}

\end{document}